\documentclass[pra,aps,twocolumn,superscriptaddress,nofootinbib,longbibliography,notitlepage]{revtex4-2}

\usepackage{amsmath}  \usepackage{amssymb}  \usepackage{amsfonts}  \usepackage{bm}
\usepackage{braket}  \usepackage{color}  \usepackage{graphicx}  \usepackage{mathtools}
\usepackage{xcolor}  \usepackage{float}
\usepackage[colorlinks,linkcolor=blue,citecolor=blue,urlcolor=blue,hyperindex,driverfallback=dvipdfm]{hyperref}
\usepackage[T1]{fontenc}
\usepackage{etoolbox}
\usepackage{orcidlink}
\usepackage[normalem]{ulem}  

\def\ii{{\rm i}}  \def\ee{{\rm e}}
\def\me{m_{\rm e}}

\newcommand{\coh}{\mathrm{coh}}

\definecolor{dgreen}{rgb}{0.00,0.45,0.15}
\definecolor{litblue}{rgb}{0.00,0.20,0.85}

\begin{document}
\def\bibsection{\section*{\refname}}

\title{Synthesizing free-electron wave functions by stimulated near-field interactions}

\author{\'{A}lvaro~Rodr\'{i}guez~Echarri\,\orcidlink{0000-0003-4634-985X}}
\email[Corresponding author: ]{a.rodriguezecharri@amolf.nl}
\affiliation{Center for Nanophotonics, NWO Institute AMOLF, 1098 XG Amsterdam, The Netherlands}

\author{Albert~Polman\,\orcidlink{0000-0002-0685-3886}}
\affiliation{Center for Nanophotonics, NWO Institute AMOLF, 1098 XG Amsterdam, The Netherlands}

\date{\today}

\begin{abstract}
Stimulated electron-optical-near-field interactions imprint a coherent, phase-coherent modulation on the wave function of a free electron, of which the free-space dispersion subsequently converts into a train of attosecond density peaks. Light thereby becomes a tool for synthesizing the electron wave function itself, setting when, where, and how narrowly the electron density concentrates. Here we study photon-induced near-field electron microscopy (PINEM) to gain control over that synthesis in two stages: first for a single PINEM interaction, whose design parameters we obtain in closed form, and then for multiple PINEM interactions acting in parallel, which extends the control across space and time. The design rules follow from decomposing the propagated density into temporal harmonics, which we classify into three:  (i) the arrival time of the attosecond train, fixed by the optical phase of the coupling; (ii) the distance at which the train forms, set by diffraction at the point where all classical trajectories converge; and  (iii) the pulse duration, fixed by the imprinted energy spread.  We evaluate these parameters for a scanning-electron-microscope (SEM) configuration at 10~keV, where the compression completes within tens of micrometers from the interaction region, with results 20\% better than previous approaches. Letting the electron then interact with several spatially separated near fields in parallel, normal to the trajectory, each pathway carrying its own coupling strength and optical phase, we show that retaining or erasing which-path information changes both the propagated density and the measurable energy spectrum, the latter providing in turn a measure of the mutual coherence of the driving fields. These results establish PINEM as a quantitative synthesis tool for free-electron wave functions in compact, low-voltage electron microscopes.
\end{abstract}

\maketitle

\section{Introduction}
Free electrons are among the most versatile probes of matter: their picometre de
Broglie wavelengths resolve atomic structure, and their femto- to attosecond
temporal structure can follow ultrafast electronic and optical dynamics. Both capabilities
derive from the electron wave function, and a central
goal of modern electron microscopy is to gain the same degree of control over
that wave function that optics has long exercised over light. Optical fields have
emerged as the natural tool for this task, and the available mechanisms differ in
which regime of the electron-photon interaction they are in. Elastic schemes exploit the
ponderomotive $\mathbf A^2$ coupling: free-space modulators driven by spatially
shaped light act on the transverse profile of the wave function, and therefore on
the profile of the beam, while preserving the electron energy for
spectrally narrow illumination \cite{garciadeabajo_prl_2021};
programmable shaping of this kind has been demonstrated with a spatial light
modulator in a $30$-keV ultrafast scanning electron microscope, realizing
electron lenses of millimetre-scale focal length
\cite{chiritamihaila_prx_2022}. A two-colour ponderomotive travelling
wave in vacuum instead modulates the wave function longitudinally, compressing
$23.5$~keV electrons into trains whose individual pulses are shorter than
$300$~as \cite{kozak_prl_2018,
kozak_prl_2019}. Further elastic schemes include Kapitza--Dirac diffraction from an optical
standing wave, used in a scanning microscope to measure electron pulse durations
in situ \cite{moriova_rsi_2025}, fields abruptly truncated at a mirror
\cite{kirchner_natphoton_2014}, and single-cycle terahertz compressors, which
place the temporal focus of an isolated electron pulse at a chosen plane
\cite{kuttruff_sciadv_2024}. Incidentally, inelastic schemes exchange energy with a
confined near field through a stimulated process, where the electron absorbs
photons from, and emits photons into, the externally driven field, so the
exchange is coherent and phase locked to the drive. Laterally raster-scanning the electron beam creates photon-induced
near-field electron microscopy (PINEM)
\cite{barwick_nature_2009,garciadeabajo_nanolett_2010,
park_njp_2010,feist_nature_2015,polman_natmater_2019}: an
electron traversing the near field of an illuminated nanostructure exchanges many
photons, and its spectrum develops a comb of sidebands spaced by the driving photon energy $\hbar\omega$, with the comb structure a measure of the near-field intensity.
Combining structured near fields with sequential interactions extends this
control to the full spatio-temporal domain, in which transversely imprinted phases have
produced electron vortex beams and holograms
\cite{vanacore_natmater_2019,madan_sciadv_2019}, and
phase-locked interaction sequences give attosecond temporal control
\cite{vanacore_natcommun_2018,vanacore_rivnuovocim_2020,
ruimy_natphys_2025}.

All those recent developments have put the shaping of the electron wave function at the focus of research \cite{vanacore_rivnuovocim_2020,garciadeabajo_acsphotonics_2021,ruimy_natphys_2025},
for which a general framework has been formulated
\cite{reinhardt_acsphotonics_2020}, later extended to light of arbitrary photon
statistics \cite{digiulio_optica_2020} and to matter waves in general
\cite{akbari_sciadv_2022}, and used to propose qubits encoded in the energy ladder of the
electron \cite{reinhardt_annphys_2020}. In the transverse direction,
the pattern of a plasmonic standing wave
can be tailored so that the electron acquires a tunable lateral profile, Bessel
beams and vortex arrays among them, with the transverse coherence of the imprint
confirmed by letting the modulated beam diffract
\cite{tsesses_natmater_2023}. Longitudinally, Ref.~\cite{lastovickovastreshkova_prresearch_2026} showed that
light engineered in time makes the sideband spacing and populations
programmable in time, imprinting on one sideband a chirp that free propagation
undoes, so that $5$--$30$~keV electrons stretched to hundreds of femtoseconds
recompress to about ten femtoseconds in that filtered sideband. 
In this work, we build on the progress achieved so far, where the drive stays quasi-monochromatic, 
so that each interaction is specified by the coupling parameter between the near field and the electron, thereby shaping its wave function.

The sideband comb is the spectral fingerprint of a phase imprinted on the electron,
and that phase, rather than the comb, is the quantity that makes the electron wave function evolve in time.
The interaction is coherent and reversible, as established by the
observation of multilevel Rabi oscillations among the sidebands
\cite{feist_nature_2015}, and it leaves the electron density unchanged at
the exit of the interaction region. The structured wavefunction appears only during subsequent
free propagation, where the quadratic term of the electron dispersion converts
the sinusoidal phase into a train of density peaks spaced by the optical period,
in direct analogy with the Talbot self-imaging of a periodically modulated
optical field \cite{talbot_philmag_1836,
rayleigh_philmag_1881}, an analogy that has
been derived and probed experimentally for a laser-modulated electron beam \cite{tsarev_prresearch_2021}. Attosecond electron pulse trains produced in this
way have been generated and characterized in transmission electron microscopes
\cite{priebe_natphoton_2017,vanacore_natcommun_2018}, extended to sub-cycle
microscopy of optical near fields \cite{nabben_nature_2023}, and exploited for
homodyne detection of field dynamics \cite{gaida_natphoton_2024}. Additionally, trains
obtained by the closely related route of modulation in the optical field cycles
at a dielectric membrane have delivered atomic-resolution diffraction and real-space imaging
\cite{morimoto_natphys_2018}.
There are also examples with continuous-wave illumination \cite{liu_acsphotonics_2019,
das_ultramic_2019} instead of pulsed ones, such as interaction driven in integrated photonic cavities
\cite{henke_nature_2021,kfir_nature_2020},
and quantum light of tailored photon statistics
\cite{dahan_science_2021}.
In an alternative experiment, photons generated by the electron itself in
a nanostructured emitter can re-excite the specimen downstream, and the resulting
cathodoluminescence interference reveals up to $27\%$ mutual coherence between
the electron-induced and photon-induced polarizations in a scanning electron
microscope \cite{taleb_natphys_2023}.

Almost all of the work above has been carried out in transmission electron
microscopes at $60$--$300$~keV electron energies. Scanning electron microscopes
operate at much lower energies, a regime in
which the modulation of nonrelativistic electrons by an optical near field was
studied well before PINEM \cite{ishikawa_journalofappli_2001}; however, quantum-coherent PINEM has been
demonstrated in an SEM down to $10.4$~keV, with up to twelve photon orders exchanged at
$17.4$~keV \cite{shiloh_prl_2022}.
These experiments characterized the interaction through energy spectra and images, whereas they did not directly measure or design the attosecond density structure of the wave function produced by the interaction, in contrast to the focus of our work.
The low-energy regime is particularly well
suited to that synthesis, since the characteristic distances are reduced commensurate with SEM setups. 
The longitudinal Talbot distance scales as
$z_T \propto v^3$, so it reduces by more than an order of magnitude between $120$~keV
and $10$~keV, and the propagation over which the attosecond structure forms
shrinks from millimeters to tens of micrometers, hence the entire synthesis then takes
place within the distance between sample stage and electron column, which in the SEM is $\sim10$~mm.
The compression is moreover achieved by free propagation alone, requiring no
lens or other electron-optical element between the interaction and the focus,
so the aberrations that such elements would introduce are absent by
construction. Lower electron energies can enhance coupling to strongly confined optical modes when the electron velocity and near-field spatial profile are suitably matched \cite{digiulio_acsnano_2024,akerboom_acsnano_2024}. At $10$~keV the non-recoil approximation underlying the sideband ladder still
holds, where the genuinely quantum corrections that arise once this ceases have
been examined elsewhere \cite{synanidis_sciadv_2024}, and the recoil-resolved
ladder itself has been proposed as a programmable qudit
\cite{sirotin_arxiv_2024}. 
Yet the design rules for controlling electron wavepackets at low energy are only partly settled. The
compression distance is known to leading order \cite{baum_jap_2017} as well as the revival scale given by the Talbot distance \cite{tsarev_prresearch_2021}; what is missing is the correction that the wave nature of the
electron imposes on propagation, and a closed-form expression of both the arrival time and the resulting temporal compression.

In this work, we theoretically study the fundamental properties
of the stimulated electron-photon interaction in PINEM to design
the resulting electron wavefunction on demand.
The paper builds this capability in two steps. 
First, we identify and quantify the degrees of control offered by 
a single PINEM interaction. We show that the propagated density admits a closed-form decomposition into
temporal harmonics whose phases do not depend on the propagation distance.
All harmonics therefore come into phase at the same retarded time, which is
fixed solely by the optical phase of the coupling $g$ and sets \emph{when} the
attosecond train arrives. From the same expansion we obtain
\emph{where} the train forms. Its leading term reproduces the classical
bunching distance $z_T/(8\pi|g|)$, which for optical-cycle velocity modulation
followed by free propagation was given by Baum \cite{baum_jap_2017} and later
recast for a sinusoidally phase-modulated beam by Yalunin, Feist and Ropers
\cite{yalunin_prresearch_2021}. What that first-order estimate leaves open is the
diffractive correction. We
solve the corresponding diffraction problem and obtain a correction that is
universal in form, giving the focus position in closed form to better than
$0.5\%$ and reproducing published numerical values that the leading term misses
by $20$--$30\%$. Together with the coupling-controlled energy spread, which sets
how short each electron pulse is, these are three independent and analytical
rules to design the electron wave function, evaluated for a $10$~keV, $1064$~nm
SEM-compatible configuration.

\begin{figure*}[htbp]
    \centering
    \includegraphics[width=\linewidth]{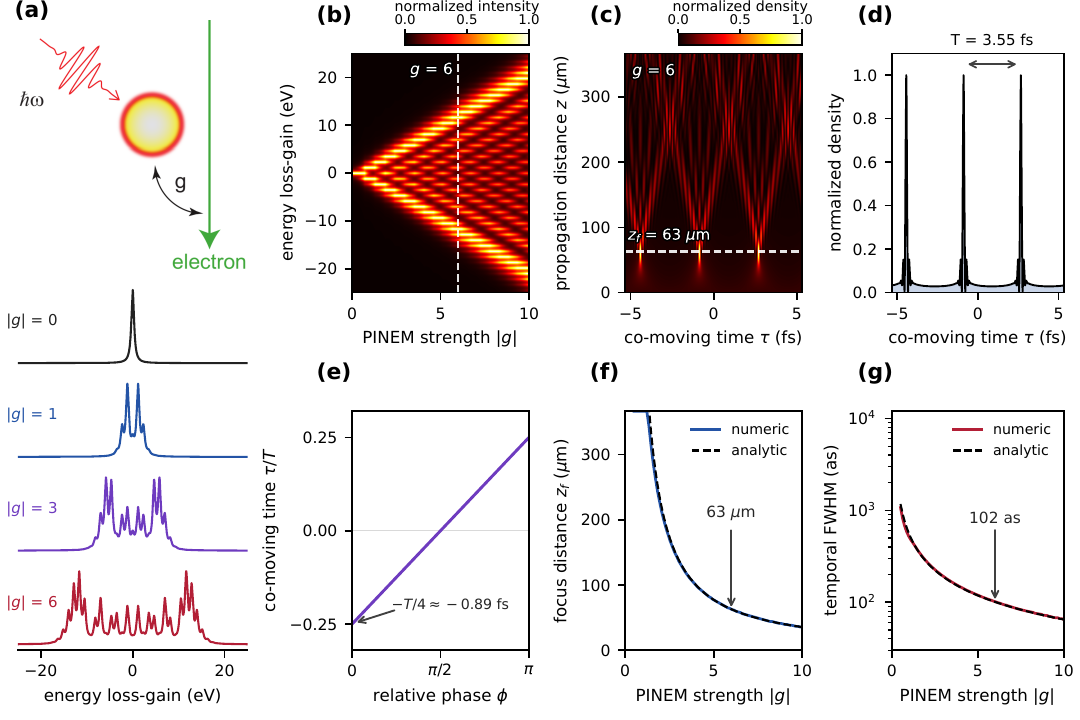}
\caption{\textbf{Synthesis of an attosecond electron wave function by a single
PINEM interaction.} All panels are computed for $10$~keV electrons and a
$1064$~nm excitation ($\hbar\omega=1.165$~eV, $T=3.55$~fs, $z_T=7.33$~mm).
\textbf{(a)} Illustration of a swift electron traversing the near field generated by an optical mode excited at energy $\hbar \omega$. The coupling parameter $g$ controls the electron-photon interaction, and the resulting normalized energy-loss spectra are shown for $|g|=0$, $1$, $3$ and $6$
(with a Lorentzian broadening of $0.7$~eV at full width at half maximum).
\textbf{(b)} Energy-loss spectrum as a function of coupling strength, with the
same $0.7$~eV broadening, each column normalized to its maximum. The comb
broadens with increasing coupling $|g|$, its outermost sidebands following
$\pm2|g|\hbar\omega$ ($\pm14$~eV at $|g|=6$, dashed white line).
\textbf{(c)} Propagated charge density $\rho(z,\tau)$ for $|g|=6$ in the
co-moving frame, normalized to the incident density. The white dashed line marks
the analytic focus of Eq.~\eqref{eq:zf_practical_main}, $z_f\approx63.0~\mu$m.
\textbf{(d)} Temporal cut at the focus: a train of attosecond peaks separated by
the optical period $T=3.55$~fs.
\textbf{(e)} Retarded time position of the focus versus the optical phase of the
coupling.
\textbf{(f)} Focus distance as a function of the coupling strength: numerical propagation
(solid) and the closed-form result of Eq.~\eqref{eq:zf_practical_main} (dashed).
\textbf{(g)} Temporal full-width at half maximum (FWHM) at the focus versus coupling
strength (solid), compared with the closed form of
Eq.~\eqref{eq:fwhm_main} (dashed). The arrow marks the $\sim102$ attoseconds (as) reached at
$|g|=6$.
}
\label{fig1}
\end{figure*}

In the second part of the paper, we turn to multi-PINEM geometries, which
add another degree of freedom: the spatial
pathway. There have been sequential schemes, in which one electron traverses two or
more interaction regions in series creating Ramsey-type phase control
\cite{echternkamp_natphys_2016}, multi-zone momentum-state preparation
\cite{feist_prresearch_2020}, and free-electron homodyne
detection \cite{gaida_natphoton_2024}; there the electron stays in a
single spatial mode and the couplings add at the amplitude level automatically.
We consider instead the parallel configuration, in which distinct
components of one wave function traverse different, non-overlapping near fields
and are subsequently recombined. Transversely structured coupling has been
imprinted and suggested in several geometries
\cite{garciadeabajo_prb_2016,cai_prb_2018,
tsesses_natmater_2023}, and parallel interactions across a shaped
interaction plane have been proposed as a route to simultaneous spatial and
temporal focusing \cite{garciadeabajo_prl_2023}, namely controlling the compression characteristics of the wavefunction. Here, the objective shifts from controlling the electron’s spatial focusing to directly designing its wavefunction. 
Each pathway interacts with an independently controlled field, characterized by its own interaction strength and optical phase, allowing different components of the wavefunction to be manipulated separately. 
The final state emerges upon recombination of these pathways. Whether they can be recombined into a single coherent wavefunction depends on whether the apparatus or environment retains information about which path was taken. If such which-path information is available, the pathways combine incoherently, as classical probabilities corresponding to independent beams. Only when this information is absent or erased can the pathways interfere and reconstruct a coherent wavefunction \cite{henke_pra_2025,kfir_sciadv_2021}.

Multi-PINEM thus provides a route toward full spatiotemporal control of the electron wavefunction. By assigning a different interaction to each pathway, independent control can be extended across both transverse dimensions and time. At the same time, the energy spectrum after recombination reflects the mutual coherence of the driving fields, providing a form of coherence spectroscopy in which the electron itself serves as the interferometric probe.

\section{Synthesizing the wave function with a single PINEM interaction}
\label{sec:single}

In a PINEM interaction, an electron prepared in a well-defined energy
state and coupled to a monochromatic near field
of angular frequency $\omega$, the interaction is described by a displacement
operator on the energy ladder $|N\rangle$, and the state immediately after the
interaction region reads $|\psi_{\rm out}\rangle=\sum_N c_N(g)|N\rangle$ with
\begin{equation}
c_N(g)=\left(\frac{g}{|g|}\right)^{\!N} J_N\!\left(2|g|\right),
\qquad
P_N=J_N^2\!\left(2|g|\right),
\label{eq:cN_main}
\end{equation}
where $P_N=|c_N|^2$ is the probability of finding the electron
in the $N$th sideband (see Methods), and
the single dimensionless complex number
$g=(e/2\hbar\omega)\int dz\,E_z(z,\omega)\,\ee^{-\ii(\omega/v)z}$, the so-called coupling parameter, which encodes the field and geometry information through the phase-matched Fourier component of the longitudinal field along the electron trajectory. Therefore, we have that $2\hbar\omega\Re\{g\}$ measures the
effective optical work the field can perform on the electron, and $\phi = \arg\{g\}$
carries the optical phase of the drive
\cite{garciadeabajo_njp_2008,garciadeabajo_nanolett_2010,digiulio_optica_2019,feist_nature_2015}. Figure~\ref{fig1}(a)
sketches the geometry, where an electron passes through the near field of an
illuminated nanostructure (as an example of a mechanism to create a near-field distribution), and the sideband combs in the lower part show how the spectrum
develops as the coupling $|g|$ increases. In the time domain, this corresponds to the fact that the wave function acquires the sinusoidal phase factor
$\exp[-2\ii|g|\sin(\omega\tau-\arg\{g\})]$, where $\tau=t-z/v$ is the retarded time
in the frame co-moving with the electron. The charge density leaves the interaction region unchanged, while the pulsed structure emerges only upon propagation, as the acquired phase modulation disperses into a density modulation. The resulting attosecond pulse train, including its position and duration, is therefore determined by the phase imprinted during the interaction.

\begin{figure}
\centering
\includegraphics[width=\linewidth]{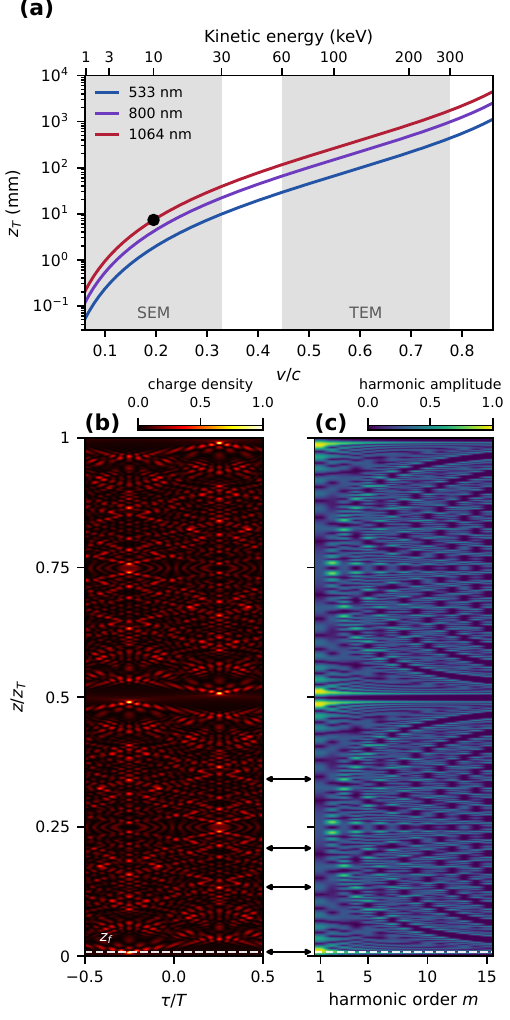}
\caption{\textbf{Propagation of the electron wavefunction.}
\textbf{(a)} Talbot distance versus electron velocity for three common drive
wavelengths; the upper axis gives the corresponding kinetic energy.
The shaded bands mark the ranges over which an SEM ($1$--$30$~keV) and TEM ($60$--$300$~keV) operate. The marker
denotes the working point of this work, $10$~keV and $1064$~nm, where
$z_T=7.33$~mm.
\textbf{(b)} Propagated charge density over one full Talbot period for $|g|=6$,
in normalized coordinates.
\textbf{(c)} Magnitude of the temporal harmonics $|\mathcal I_m(z)|$ of the same
carpet, from Eq.~\eqref{eq:harmonics_main}, sharing the propagation axis with
(b); only positive orders are shown. The dashed line marks the attosecond focus at $z_f\approx63~\mu$m (i.e. $z_f/z_T \approx 0.01$), same as in Fig.~\ref{fig1}(c). Black arrows indicate the plane cuts of the charge density and the harmonic amplitude shown in Fig.~S5 in the SI.}
\label{fig2}
\end{figure}

Figure~\ref{fig1} collects the resulting synthesis for the benchmark system of
this work: electrons of kinetic energy $E_0=10$~keV ($v=0.195c$, Lorentz factor
$\gamma=1.020$) driven at $\lambda=1064$~nm ($\hbar\omega=1.165$~eV).
Figure~\ref{fig1}(b) shows the measured signature of the phase
imprint, namely the energy loss-gain spectrum convolved with a Lorentzian line shape
of $0.7$~eV full width, representing a realistic instrumental resolution. As
$|g|$ grows, the comb extends out to the characteristic maxima of the Bessel
amplitudes at $\pm2|g|\hbar\omega$, reaching $\pm 14$~eV at $|g|=6$.
Figure~\ref{fig1}(c) shows what the same state does upon free propagation. The quadratic term of the electron dispersion multiplies each
sideband by $\exp(-2\pi\ii N^2 z/z_T)$, with the longitudinal Talbot distance
(see Methods)
\begin{equation}
z_T=\frac{4\pi\me v^3 \gamma^3}{\hbar\omega^2},
\label{eq:zT_main}
\end{equation}
and the phase modulation converts into a train of attosecond (as) pulses
separated by the optical period $T=2\pi/\omega=3.55$~fs. At $|g|=6$ the train
forms after only $z_f\approx63~\mu$m and each pulse is $\sim100$~as long, as observed in Fig.~\ref{fig1}(d). We plot in Fig.~\ref{fig2}(a) the values of the Talbot distance for different excitation wavelengths as a function of the electron speed, which in our setup results in $z_T \approx 7.33$~mm (see black dot). That is two orders of magnitude smaller than if we operate with 200~keV electrons in a TEM ($z_T \approx 845$~mm), making low-energy SEM chamber sizes very well suited to resolving the intricate diffraction pattern of the electron wave function.

In paraxial optics, the Talbot distance is defined when a periodic modulation of
period $a$ imprinted on a wave of wavelength $\lambda$ self-images after a distance $ \approx2a^2/\lambda$ for $\lambda\ll a$
\cite{talbot_philmag_1836,rayleigh_philmag_1881}.
Analyzing the PINEM-modulated electron as a matter wave of de Broglie
wavelength $\lambda_B=2\pi\hbar/(\gamma\me v) \approx 12.2$~pm at $10$~keV, and
using Eq.~\eqref{eq:zT_main} with the corresponding optical counterpart, we obtain an effective
longitudinal period
\begin{equation}
a\approx\sqrt{\frac{z_T\lambda_B}{2}}=\gamma v\,\frac{2\pi}{\omega}=\gamma v T,
\label{eq:talbot_period_main}
\end{equation}
which is the distance the electron travels in one optical cycle, yielding $\sim 212$~nm in our setup. The longitudinal Talbot effect is therefore the exact
energy--time counterpart of near-field diffraction from a grating of that period:
the optical interaction writes a periodic phase along the beam, and free
propagation converts it into a density modulation that self-images at $z_T$ and
half-images at $z_T/2$. The same revival scale has been derived for a laser-modulated electron beam
by Tsarev, Ryabov and Baum, who place the phase-mirrored configuration at $z_T/2$
and complete restoration at $z_T$, and who report spectroscopic evidence for the
intermediate fractional revival at $z_T/4$, where the train reaches unit temporal
contrast \cite{tsarev_prresearch_2021}. Figure~\ref{fig2}(b) shows one full period of this carpet.

\subsection{Spatial and temporal characterization of the attosecond focus}
\label{subsec:focus}

To understand the spatio-temporal focusing behavior of the electron wave function, we pay close attention to the Fourier series of the Talbot carpet, plotting it for our setup in Fig.~\ref{fig2}(c). Every harmonic of the optical period oscillates along $z$ with its own spatial frequency, faster for higher orders,
much as in a pendulum wave built from pendula of staggered periods. All
harmonics rise together at one particular distance, and their constructive
addition builds the sharp attosecond pulses; further along they fall out of step
again and the carpet dissolves into its revival pattern. Therefore, the position of the observation plane selects which harmonics dominate the density modulation \cite{tsarev_prresearch_2021}. By choosing the appropriate position, these harmonics could be matched to resonant modes of a microcavity or plasmonic particle, enabling the electron to probe the corresponding optical modes. For the sake of better visualization, we plot in Fig.~S5 in the supplementary information (SI), different cuts of the Talbot carpet at the positions indicated by the black arrows in Fig.~\ref{fig2}b,c.

\textbf{When the attosecond train arrives.} Expanding the propagated density in temporal
harmonics of the optical period, $n(z,\tau)=\sum_m \mathcal
I_m(z)\,\ee^{-\ii m\omega\tau}$, where $n$ is the longitudinal
electron density referred to its incident mean, so that the charge density is
$-en$, inserting Eq.~\eqref{eq:cN_main} into the
propagated wave function, and carrying out the sum over sidebands yields the closed form (see Methods)
\begin{equation}
\mathcal I_m(z)
=
\ee^{\ii m(\phi-\pi/2)}\,
J_m\!\left(4|g|\sin\frac{2\pi m z}{z_T}\right),
\label{eq:harmonics_main}
\end{equation}
with $\phi=\arg\{g\}$. The Bessel coefficient is real but can be positive or negative, so a
harmonic whose argument or whose Bessel value turns negative carries an extra phase of
$\pi$. Both happen only beyond the first Bessel zero, so in the region where the
attosecond peak forms every coefficient is positive. The magnitude of Eq.~\eqref{eq:harmonics_main} is known: the
same Bessel envelope appears as the $m$-th density moment of a sinusoidally
phase-modulated beam in Ref.~\cite{yalunin_prresearch_2021} and, squared,
as the degree of coherence used to quantify attosecond bunching in Ref.~\cite{garciadeabajo_prl_2023}; the same quantity has also been written as an explicit sum over products of Bessel functions, without being carried out in closed form \cite{kfir_sciadv_2021}. Importantly, the phase in Eq.~\eqref{eq:harmonics_main} of
every harmonic is independent of the propagation distance and strictly linear in
$m$, up to those sign changes. Therefore all harmonics are simultaneously in phase at the retarded times
\begin{equation}
\omega\tau_f=\phi-\frac{\pi}{2}+2\pi\ell,
\qquad \ell=0,\pm1,\pm2,\dots
\label{eq:tau_star_main}
\end{equation}
These times $\tau_f$, written by analogy
with the focal distance $z_f$, are where the attosecond peaks form, and free
propagation only redistributes the harmonic amplitudes.
The temporal position of the train is thus set by the optical phase alone, independently of the coupling strength and of the
propagation distance. We plot the resulting linear dependence of $\tau_f$ in Fig.~\ref{fig1}(e) with respect to the coupling phase $\phi = \arg{\{g\}}$.

\textbf{Where the attosecond train forms.} In the pre-revival region ($z\ll z_T$) the amplitudes of each harmonic of the Fourier series peak at nearly the same distance, which defines the focal plane. In this regime, the linearization of the periodic phase $\sin(2\pi mz/z_T)\simeq 2\pi mz/z_T$ reduces the amplitudes to
the classical bunching form $J_m(mu)$ of velocity-modulated beams \cite{kutsaev2021electron,pan2023weak}, with the
scaled coordinate $u=8\pi|g|z/z_T$ shared by every harmonic. Because the first
maximum of $J_m$, at $j'_{m,1}=m+0.8086\,m^{1/3}+\mathcal O(m^{-1/3})$
[Eq.~(9.5.16) in Ref.~\citenum{AS1972}], approaches $m$ at large order, all
harmonics peak in unison at $u\simeq1$, giving $z_f\simeq z_T/(8\pi|g|)$. 
This is the classical compression distance. Baum \cite{baum_jap_2017} writes it in
Eq.~(1) as $L_{\rm focus}=v^2T/(2\pi\Delta v_{\max})$ for a generic peak velocity
modulation $\Delta v_{\max}$. For PINEM the comb edge is
$2|g|\hbar\omega$ and the longitudinal effective mass of Eq.~\eqref{eq:zT_main} is
$\gamma^3\me$, so $\Delta v_{\max}=2|g|\hbar\omega/(\gamma^3\me v)$, which returns
$z_T/(8\pi|g|)$. Yalunin, Feist and Ropers give the same result in Eq.~(7) of Ref.~\cite{yalunin_prresearch_2021}, obtained in the classical limit as the
paraxial temporal focus of the ballistic map of a sinusoidally phase-modulated
beam.

However, in this approximation, the analytical expression yields a shorter distance than that obtained by numerical simulation (see Fig.~S2 in the SI). Some of the approximations made at $u=1$ (i.e., where all classical trajectories arrive at the same time) require higher-order terms to improve the results. Incidentally, different harmonics are $m^{1/3}$ out of phase, so small and high harmonics do not peak at the same position. Figure~S1 of the SI resolves that spread around the focus. The true maximum is therefore set by diffraction instead of by trajectory crossing. 
Retaining the higher-order terms of the Bessel phase reduces the problem to a standard diffraction integral
-- the on-axis Pearcey function that governs wave fields near a focus with
spherical aberration -- whose maximum occurs at a universal value of a single
scaled detuning (see Methods and Sec.~S3 of the SI). The result
is
\begin{equation}
z_f=\frac{z_T}{8\pi|g|}\, \left[1+\frac{0.898}{\sqrt{2|g|}}+\frac{0.44}{2|g|}
+\mathcal O\!\left[(2|g|)^{-3/2}\right] \right]
\label{eq:zf_practical_main},
\end{equation}
where the leading correction coefficient $0.898$ is derived from the
Pearcey maximum and the second is fitted to the exact harmonic sum.
Equation~\eqref{eq:zf_practical_main} reproduces the exact focus to better than
$0.5\%$ for $|g|\geq4$ and is the curve overlaid on the numerical data in
Fig.~\ref{fig1}(f), where, for instance, the classical estimate would underestimate $z_f$ by
$23\%$ at $|g|=6$ (see also Fig.~S3 of the SI, which follows the density at
$\tau_f$ along the focus for $|g|=5$, $6$ and $7$).

In the literature, we find examples where Eq.~\eqref{eq:zf_practical_main} above can be applied. In particular, two cases in which the PINEM scenario has been studied under different setups. For $120$~keV
electrons at $800$~nm, Feist \emph{et al.} \cite{feist_nature_2015}
simulate optimal temporal compression $1.8$~mm behind the sample at $|g|=5.7$;
at those parameters Eq.~\eqref{eq:zf_practical_main} gives $1.83$~mm, while the
classical term alone gives $1.40$~mm. Priebe \emph{et al.}
\cite{priebe_natphoton_2017} predict the shortest pulses after
$2.75$~mm of propagation at $|g|=3.95$ (their experiment was geometrically
limited to $1.5$~mm, short of the focus, where $655$~as peaks were measured);
here Eq.~\eqref{eq:zf_practical_main} gives $2.78$~mm and the classical term
$2.02$~mm (see  Fig.~S6). The closed form therefore recovers both simulated values to within
$2\%$.

\textbf{How short are the pulses?}
The interaction imprints the phase factor
$\ee^{-2\ii |g|\sin(\omega\tau-\phi)}$ (see Eq.~(S19) of Sec.~S3 of the SI). Expanding it about
the focus time, where $\omega\tau_f=\phi-\pi/2$, gives
$2|g|-|g|\,x^2+|g|\,x^4/12-\dots$ with $x=\omega(\tau-\tau_f)$. The
quadratic term is the chirp that the quadratic dispersion of free propagation
cancels at $z_f$, and the constant is irrelevant. However, the surviving quartic term produces the temporal analogue of spherical aberration, replacing the ideal point focus with a cusp caustic and preventing the pulse from reaching its transform limit.
A single
interaction produces a standard deviation of the population number of $\sigma_N=\sqrt2\,|g|$ in units of the
sideband index $N$, that is,
$\sigma_E=\sqrt2|g|\hbar\omega=9.9$~eV at $|g|=6$, an
equivalent transform-limited Gaussian full width at half maximum 
$\Delta E_{\mathrm{TL}}=2\sqrt{2\ln2}\,\sigma_E=4\sqrt{\ln 2}\,|g|\hbar\omega
=23.3~\mathrm{eV}$, to be compared with the $4|g|\hbar\omega=28$~eV that separate the outer Bessel
maxima of the comb, and 26.3~eV obtained from the numerical propagation. However, as we discuss in Sec.~S4 of the SI, the generated attosecond pulses do not follow any standard lineshape.
The assumption that the spectral width 
terminates at the edges of the $P_N$ distribution, overestimates the value of $\Delta E$  by a correcting factor of the order of $|g|^{1/3}$, which corrects that estimate to
\begin{equation}
    \Delta E=4\hbar\omega|g|\left[1-\frac{0.0729}{(2|g|)^{2/3}}\right],
\label{eq:spec_fwhm_main}
\end{equation}
giving $\Delta E = 27.6$~eV at the working setup. Incidentally, the temporal width follows from the same peak-diffraction analysis that
fixes $z_f$, derived from the imprinted phase, as
\begin{equation}
\Delta\tau=\frac{0.888}{\omega\,(2|g|)^{3/4}}
\left[1+\frac{0.898}{\sqrt{2|g|}}+\frac{0.44}{2|g|}\right]
\label{eq:fwhm_main}
\end{equation}
that reproduces the numerical width better
than $2\%$ over $2\le|g|\le30$, giving $100.9$~as against $101.6$~as at the
working point ($|g|=6$), see Fig.~\ref{fig1}(g), which is better than assuming a Gaussian time-bandwidth product that yields a 78.4~as pulse length; Fig.~S4 of the SI collects the spectral width, the duration and their product against the coupling strength.

\subsection{Longitudinal scaling at low electron energy}
\label{subsec:lowE}

Couplings far exceeding unity are routine at transmission-microscope energies,
reaching $|g|\simeq150$ at $207$~keV by resonant inverse-Cherenkov phase matching
\cite{dahan_natphys_2020} and $|g|\simeq125$ on a $120$~keV beam from tens of
milliwatts of continuous-wave power in a high-$Q$ Si$_3$N$_4$ microresonator
\cite{henke_nature_2021}. Geometry and drive nevertheless matter more than beam
energy, as continuous-wave illumination of a bare $50$-nm silicon membrane yields
only a fitted $|g|\simeq0.3$ at $75$~keV \cite{tsarev_prresearch_2021}. At
scanning-microscope energies, the benchmark adopted here, up to twelve photon
orders are exchanged per side at $17.4$~keV \cite{shiloh_prl_2022}, which implies
$|g|\simeq6$ through the $2|g|$ cutoff, although that work quotes no coupling for
its spectra. Combs of comparable depth have been resolved at $25$~keV in the truncated field
of a metal mirror, ten photon orders on the gain side spaced by the $1.55$~eV
photon energy \cite{kirchner_natphoton_2014}. Other experiments at these same
beam energies transfer thousands of photons at once, so their spectra show a
broad gain tail instead of separate orders
\cite{kozak_jap_2018,chlouba_nature_2023}.

As the longitudinal scale follows $z_T\propto\gamma^3v^3\lambda^2$,
lowering the electron energy from $200$~keV to $10$~keV shortens it by two
orders of magnitude at $1064$~nm drive, from $z_T\approx845$~mm to
$z_T\approx7.33$~mm, at a coupling strength $|g|=6$ that is a conservative value.
The shaded bands of Fig.~\ref{fig2}(a) mark where the two electron microscopy classes operate. At fixed $|g|=6$, the focus of Eq.~\eqref{eq:zf_practical_main} stays below $0.34$~mm anywhere in the
SEM electron energy range and is $1$~mm or more in the TEM.
However, lowering the
electron energy in fact strengthens the
attainable coupling. 
For instance, in
the quasistatic limit the coupling probability of a three-dimensional
nanostructure scales as $1/v$ at fixed interaction phase $\omega D/v$, and for a two-dimensional polaritonic film with an intrinsic resonance, where the scaling is $1/v^2$ \cite{digiulio_acsnano_2024}. 
At $10$~keV and
$1064$~nm the phase-matching condition fixes a natural interaction length
$\pi v/\omega\simeq104$~nm for a single structure, or a field-modulation period
of $207$~nm for an extended one, with an evanescent reach
$\gamma v/\omega\simeq34$~nm setting the beam--surface distance.

Because the carpet of Fig.~\ref{fig2}(b) is universal in the scaled
coordinates $(z/z_T,\tau/T)$, which is the content of
Eq.~\eqref{eq:harmonics_main}, the focus distance factorizes into a function of
$|g|$ times $z_T(E_0,\lambda)$. The single map $z_f(|g|,z_T)$ therefore
carries the complete parameter dependence, and the maps against electron energy
or drive wavelength follow from it by evaluating $z_T(E_0,\lambda)$. We give all
three over a broad parameter range, with isocurves of constant $z_f$, in
Fig.~S6 of the SI.

\begin{figure*} 
\centering
\includegraphics[width=\linewidth]{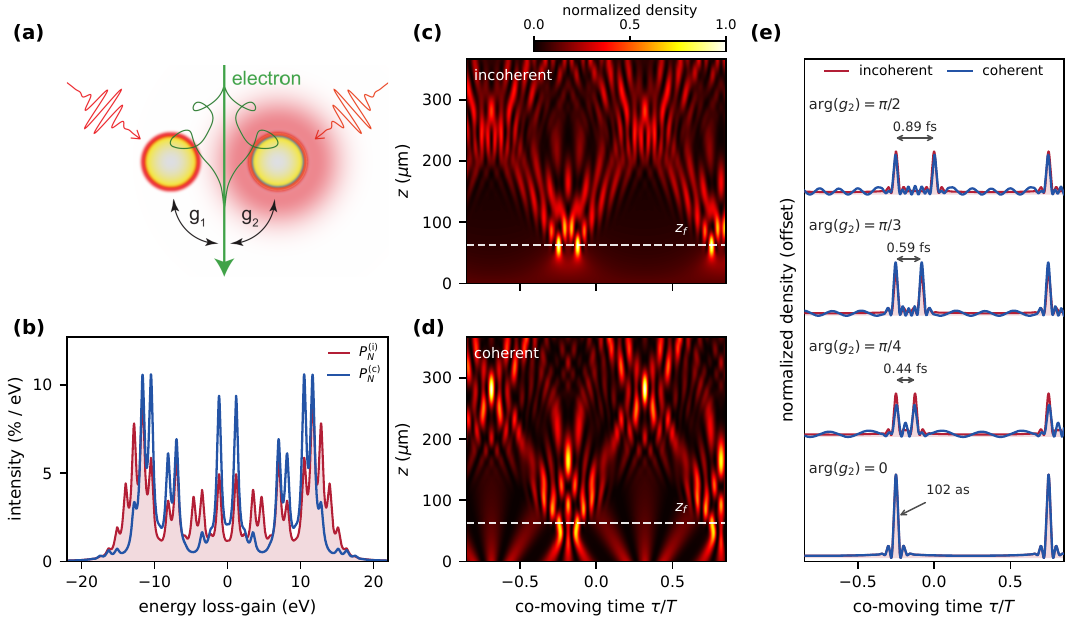}
\caption{\textbf{Two-path PINEM and phase control of the synthesized electron pulse.}
\textbf{(a)} Illustration of an electron passing between two
independently illuminated near fields, acquiring couplings $g_1$ and $g_2$.
Throughout the figure the wave function is split evenly between
the two arms, $p_1=p_2=1/\sqrt2$, so that the incoherent case of
Eq.~\eqref{eq:mixture_main}, with $\Gamma_{ij}=|p_i|^2\delta_{ij}=\delta_{ij}/2$,
is drawn in red, and the coherent, which-path-erased case of
Eq.~\eqref{eq:coherent_main}, with $\Gamma_{ij}=p_ip_j^*=1/2$, in blue.
\textbf{(b)} 
Sideband spectra for equal coupling strengths
$|g_1|=|g_2|=6$ and a relative phase $\arg\{g_2\}=\pi/4$, broadened by the
$0.7$~eV Lorentzian instrumental response, in per cent per eV.
\textbf{(c,d)} Propagated charge density for the case in (b), sharing one colour scale. The dashed line marks the position of the focus $z_f=63.0~\mu$m.
\textbf{(e)} Normalized temporal cuts at the focus $z_f$ for four values of the relative phase. At zero relative phase, the arms coincide, and the two cases are
identical; for nonzero phase, the incoherent
train splits into two of approximately half weight, separated by
$\arg\{g_2\}/2\pi$ in units of the optical period, that is by $0.44$, $0.59$ and
$0.89$~fs, while each peak keeps the $102$~as FWHM.}
\label{fig3}
\end{figure*}

\section{Multi PINEM parallel interactions}
\label{sec:parallel}

Multiple electron--light interactions have been realized mostly
sequentially, by sending the same electron through two or more interaction
regions in series. In particular, two phase-locked near fields separated by a few micrometres
implement a Ramsey-type interferometer in which the couplings add at the
amplitude level, $g_{\rm tot}=g_1+\ee^{\ii\varphi}g_2$
\cite{echternkamp_natphys_2016}; three-dimensional and multi-zone modulation
prepares high-purity momentum states \cite{feist_prresearch_2020};
a phase-locked reference interaction turns the sideband spectrum into a
homodyne detector of near-field dynamics \cite{gaida_natphoton_2024}; and a
pair of continuous-wave interactions used to rotate the
amplitude and phase of adjacent sidebands, encoding a free-electron qubit in the
$\ket N = \ket{\pm1}$ orders \cite{tsarev_prresearch_2021}.
In all of these the electron remains in a single spatial mode, so the
interference is automatic and no which-path question arises.

Here we address the complementary parallel geometry (see scheme in Fig.~\ref{fig3}a), in which different
components of one electron wave function traverse different, non-overlapping near
fields.
This is the situation whenever a broad beam illuminates several
nanostructures at once, or when the wave function is deliberately split into
distinguishable paths.  
Transversely structured coupling of this kind has been
treated theoretically for standing-wave and vortex near fields
\cite{garciadeabajo_prb_2016,cai_prb_2018} and
demonstrated experimentally with spatially shaped plasmonic fields, where the
imprinted transverse phase is preserved well enough to diffract the electron
\cite{tsesses_natmater_2023}. Parallel interactions have also been
proposed as a route to simultaneous spatial and temporal focusing: García de
Abajo and Ropers \cite{garciadeabajo_prl_2023} let transversely
separated zones of a single convergent wave function acquire distinct couplings
at one interaction plane, and recombine them with the objective lens at a common
focus, thereby numerically optimizing the zone couplings to flatten the sideband comb. While their work focuses on compressing the pulse, we aim to synthesize the wave function in space and time by allowing each of the paths to be described independently, making their mutual coherence an additional degree of freedom.

A parallel geometry consequently raises the question of how the different paths are recombined. We label the path-alternatives by an index $i$ and describe each by its own scattering
operator $\hat S_i=\hat S(g_i)$, so that the coupling strength and the
optical phase may differ from arm to arm. The path degree of freedom carries its
own reduced density matrix $\Gamma$, normalized as ${\rm Tr}\,\{\Gamma\}=1$, whose
diagonal elements $\Gamma_{ii}=|p_i|^2$ are the weights of the alternatives and
whose off-diagonal elements measure their mutual coherence. Incidentally, note that $p_i$ also defines the weight of the wave function taking the $i$ path upon preparation.
The state after the parallel-path interaction is
\begin{equation}
\hat\rho
=
\sum_{i,j}
\Gamma_{ij}\,
|i\rangle\langle j|
\otimes
\hat S_i\hat\rho^{(0)}\hat S_j^\dagger,
\label{eq:rho_after_main}
\end{equation}
with $\hat\rho^{(0)} = \ket 0 \bra 0$ for an initial monochromatic state.
Everything then depends on what happens to the paths $\Gamma_{ij}=p_ip_j^*$.
If the arms leave distinguishable signatures in any other
degree of freedom, the path is marked, $\Gamma_{ij}\to|p_i|^2\delta_{ij}$, and
only the diagonal part survives, so the electron is left in the statistical
mixture.
Thereby, the interference disappears from the single-electron signal, returning only in subensembles selected by erasing the
marker \cite{henke_pra_2025}.
Then, tracing over a label that survives the recombination
removes the $i\neq j$ terms and leaves the statistical mixture
\begin{align}
\hat\rho^{(\rm i)}
&=\sum_i |p_i|^2\,\hat S_i\hat\rho^{(0)}\hat S_i^\dagger,
\label{eq:rho_mixture_main}
\\
P_N^{(\rm i)}
&=\sum_i |p_i|^2 J_N^2\!\left(2|g_i|\right),
\label{eq:mixture_main}
\end{align}
whereas erasing it retains the off-diagonal elements of $\Gamma$ and lets the
arms interfere at the amplitude level,
\begin{align}
\hat\rho^{(\rm c)}
&=
\mathcal N_\coh^2
\sum_{i,j}
p_i p_j^*\,
\hat S_i\hat\rho^{(0)}\hat S_j^\dagger,
\label{eq:rho_coherent_main}
\\
P_N^{(\rm c)}
&=
\mathcal N_\coh^2
\left|\sum_i p_i\,\ee^{\ii N\arg\{g_i\}}J_N\!\left(2|g_i|\right)\right|^2,
\label{eq:coherent_main}
\end{align}
with the sideband populations being $P_N^{(\rm x)}=\langle N|\hat\rho^{(\rm x)}|N\rangle$
for the incoherence (x=i) and coherence (x=c) path mixtures, respectively. The normalization $\mathcal N_\coh$ enforces
${\rm Tr}\{\hat\rho^{(\rm c)}\}=1$ and its closed form is given in Methods,
Eq.~\eqref{eq:Ncoh_methods}, and reduces to unity whenever $\Gamma$ is diagonal,
so that Eq.~\eqref{eq:mixture_main} carries no such factor.
The two cases are otherwise
identical in preparation, since each arm is internally PINEM-coherent and
the interference between arms is making the difference. Note that, in coherent path recombination, if the PINEM interactions $g_i$ are not phase-locked, randomizing the phase makes the recombination incoherent, with special treatment at the $N=0$ energy level.

\begin{figure*}[htbp]
    \centering
\includegraphics[width=\linewidth]{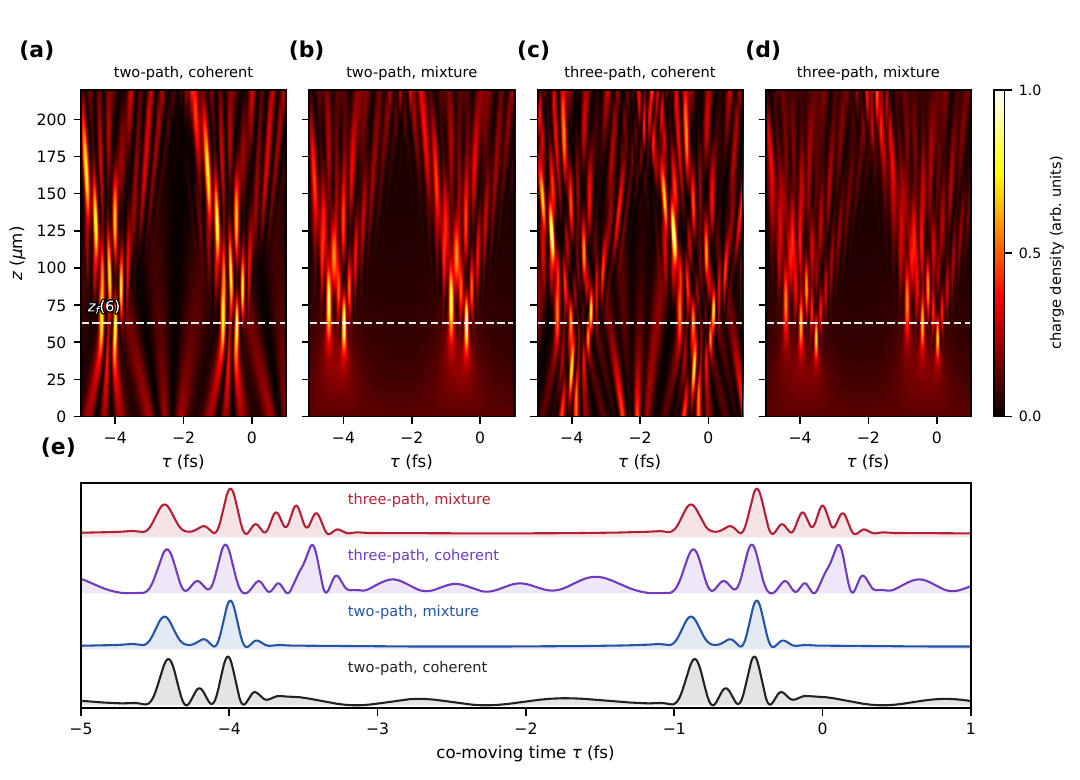}
\caption{\textbf{Propagation of parallel PINEM with unequal couplings.}
\textbf{(a--d)} Propagated charge density for two- and three-path configurations,
each in the which-path-erased (coherent) and statistical-mixture cases (incoherent). The arms carry different strengths and phases, $g_1=5$, $g_2=6\,\ee^{\ii\pi/4}$ and
$g_3=7\,\ee^{\ii\pi/2}$, so that each arm focuses at its own distance
and at its own retarded time (see table~\ref{tab:arms}). The white
dashed line marks the reference plane $z_f \approx 63.0~\mu$m at $|g|=6$.
\textbf{(e)} Temporal cuts in the $z_f(|g|=6)$ plane, each normalized to its
own maximum and offset vertically.}
\label{fig4}
\end{figure*}

\subsection{Phase control of a two-path synthesis}

Figure~\ref{fig3}(a) illustrates the scheme where an electron wave function passes between two illuminated
nanostructures (as examples of near-field sources), so that one component of it is modulated by the near field of
the first and the other by that of the second, with couplings $g_1$ and $g_2$, respectively. After the interaction, the two components then propagate
together and recombine. In particular, each arm independently synthesizes an attosecond train, and by
Eq.~\eqref{eq:tau_star_main} each train sits at its own retarded time
$\omega\tau_{f,i}=\arg\{g_i\}-\pi/2$. Provided that $|g_1|=|g_2|=6$, the two trains therefore focus at the same
distance but at different times, and the relative phase acts as a delay control parameter. Without loss of generality, we take $\arg\{g_1\}=0$ and vary $\arg\{g_2\}$ to set the desired delay, where a 10-keV electron is driven at 1064~nm. The wave function is split evenly between the arms, $\Gamma_{11}=\Gamma_{22}=1/2$, and we compare the two limits of the recombination, on one side the statistical mixture, $\Gamma_{ij}=\delta_{ij}/2$, and on the other the which-path-erased case, in which all four elements are equal, $\Gamma_{ij}=1/2$. After interaction, we show the sideband spectra in Fig.~\ref{fig3}(b). For the two scenarios, Eq.~\eqref{eq:mixture_main} returns the single-arm comb $P_N^{(\rm i)} \propto J_N^2(2|g|)$ since $|g_{1}|= |g_{2}|= 6$ , whereas Eq.~\eqref{eq:coherent_main} reduces to $P_N^{(\rm c)} \propto 2J_N^2(2|g|)\cos^2(N\arg\{g_2\}/2)$. At $\arg\{g_2\}=\pi/4$ the cosine term leads to depopulation of the sidebands $N=\pm4$ and $\pm12$ (see Fig.~S7 in the SI).

In particular, in Fig.~\ref{fig3}(c), we set $\arg\{g_2\}=\pi/4$ and compute the propagation of the charge density with an incoherent mixture of the two paths, and in Fig.~\ref{fig3}(d) the erased case under the same conditions. Thereby, we tune the peak-to-peak distance linearly with increasing $\arg\{g_2\}$, as discussed in Fig.~\ref{fig1}(e). The cuts at $z=z_f$, Fig.~\ref{fig3}(e), follow that tuning over four phases and give separations of $0.44$, $0.59$ and $0.89$~fs for $\arg\{g_2\}=\pi/4$, $\pi/3$ and $\pi/2$, with each peak $\sim 102$~as. Erasing the path label keeps the delay but not the shape, since the arms now add as amplitudes. For instance, at $\arg\{g_2\}=\pi/4$, a third burst appears midway between the other two; we also note from panel (d) that the maximum of the coherent density mixture is at a distance smaller than $z_f$ as a consequence of adding amplitudes instead of the intensities. Figure~S7 of the SI shows the two maps over a full Talbot period, together with the corresponding Fourier series. The vacancies at the harmonic orders $m=4$ and $12$ appear in the mixture and are filled by coherent recombination. This mirrors the behavior of the sidebands $N=\pm4$ and $\pm12$ in Fig.~\ref{fig3}(b), as described by $\mathcal I_m^{(\rm i)}$ and $\mathcal I_m^{(\rm c)}$, respectively. The former reduces to $|\mathcal I_m^{(\rm i)}|=|\cos(m\arg\{g_2\}/2)|\,|J_m(4|g|\sin(2\pi mz/z_T))|$, whereas the latter retains the inter-arm cross term, which has no closed form (see Methods).

\subsection{On-demand wave functions by multiple PINEM}
\label{subsec:ondemand}

The two-path example above used identical arm strengths, so a single tuning parameter was in play (i.e., the relative phase). We now exploit the possibility to tune all possible properties: arms whose
couplings differ in strength and phase, so that each pathway carries its
own complete set of single-interaction design rules from
Sec.~\ref{subsec:focus}. The intention is to show that the parallel geometry
turns those rules into individual instructions and, at the same time,
to identify the observables that reveal whether the components were recombined
coherently or merely added as classical probabilities.

As an illustrative example, Fig.~\ref{fig4} extends the analysis to arms that differ in both strength and phase. Two of them, $g_1=5$ and $g_2=6\,\ee^{\ii\pi/4}$, build the two-path case, and a third, $g_3=7\,\ee^{\ii\pi/2}$, is added for the three-path one. For an electron beam prepared at 10~keV and driven at 1064~nm, all three arms share the same Talbot distance, $z_T=7.332$~mm, as determined from Eq.~\eqref{eq:zT_main}. However, according to Eqs.~\eqref{eq:zf_practical_main} and \eqref{eq:tau_star_main}, each arm focuses at a distinct propagation distance and at a corresponding retarded time, with its own duration from Eq.~\eqref{eq:fwhm_main}, as summarized in Table~\ref{tab:arms}. 

\begin{table}[h]
\caption{Focus of each arm of Fig.~\ref{fig4}, from
Eqs.~\eqref{eq:zf_practical_main}, \eqref{eq:tau_star_main} and
\eqref{eq:fwhm_main}. The three arms share the Talbot distance
$z_T=7.332$~mm and the period $T=3.549$~fs, both fixed by the $10$~keV electron
energy and the $1064$~nm drive.}
\label{tab:arms}
\begin{ruledtabular}
\begin{tabular}{lccc}
Arm & $z_{f}$ ($\mu$m) & $\tau_{f}$ (fs) & $\Delta\tau$ (as) \\
\hline
$g_1=5$                 & $77.5$ & $-T/4=-0.887$ & $118.5$ \\
$g_2=6\,\ee^{\ii\pi/4}$ & $63.0$ & $-T/8=-0.444$ & $100.9$ \\
$g_3=7\,\ee^{\ii\pi/2}$ & $53.0$ & $0$              & $\phantom{0}88.2$ \\
\end{tabular}
\end{ruledtabular}
\end{table}

Figure~\ref{fig4}(a,b) shows the two-path configuration and
Fig.~\ref{fig4}(c,d) the three-path one, each of them with the coherent
recombination followed by the incoherent mixture. As in Fig.~\ref{fig3}, the wave
function is split evenly over the $M$ arms, with $M=2$ in (a,b) and $M=3$ in
(c,d), so that $\Gamma_{ii}=1/M$, and we compare the two limits of the
recombination, on one side the statistical mixture, $\Gamma_{ij}=\delta_{ij}/M$,
and on the other the which-path-erased case, in which all elements are
equal, $\Gamma_{ij}=1/M$. Erasing the which-path information yields additional
features, according to their interference expressions.

The cuts of Fig.~\ref{fig4}(e) are all taken in the single plane
$z_f(|g|=6)=63.0~\mu$m, and they are visibly less symmetric than those of
Fig.~\ref{fig3}(e). There, both arms carry $|g|=6$ and share one focus, so that
plane was the focal plane of both, and the arms differed only in $\arg\{g_2\}$,
which slides a train along the time axis without deforming it. Here the strengths
differ, each arm yields its own focus and
its own minimum temporal width given in Table~\ref{tab:arms}, so a single plane catches the $|g|=5$ arm short of its
focus, at $z/z_f=0.81$, and the $|g|=7$ arm past it, at $z/z_f=1.19$. 
Their different phases therefore displace
the pulses, while increasing strengths sharpen their features. Therefore, by controlling the number of paths as well as their individual properties, we can then synthesize the electron wave function on demand at different positions in time and space.

Figure~S8 of the SI shows the evolution of the charge density over a full Talbot
period. The charge density develops an intricate pattern that is determined by the
phases of the individual sidebands, and the figure presents each Talbot carpet
alongside its corresponding Fourier series. The incoherent mixture keeps only the
diagonal terms of Eq.~\eqref{eq:Im_paths_methods}, that is, the single-arm harmonics of Eq.~\eqref{eq:harmonics_main} weighted by $1/M$. For every arm the Bessel argument vanishes at $z/z_T=1/2$ and $1$, independently of the coupling strength, so that all harmonics with $m\ge1$ vanish at these planes and the mixture recovers a uniform density. The coherent case retains in addition the cross terms $i\neq j$.

The resulting spectra provide a more immediate signature of this distinction, as
shown in Fig.~S9 of the SI. The measured energy loss-gain spectrum reveals the
nature of the recombination because Eq.~\eqref{eq:coherent_main} contains
interference terms that are absent from Eq.~\eqref{eq:mixture_main}. With the
wave function split evenly over the $M$ arms, $p_i=1/\sqrt M$, the path density
matrix is diagonal in the statistical mixture, $\Gamma_{ij}=\delta_{ij}/M$, while
all elements are equal once the which-path information is erased,
$\Gamma_{ij}=p_ip_j^*=1/M$. Those off-diagonal elements let the arms add as
amplitudes, so that every pair contributes a term $\cos(N\Delta\phi_{ij})$ to the
sideband envelope, where $\Delta\phi_{ij}=\arg\{g_j\}-\arg\{g_i\}$ is the
difference between the optical phases imprinted by the two near fields. The
envelope therefore beats with a period of $2\pi/\Delta\phi_{ij}$ sidebands: the
$\pi/4$ dephasing considered here gives eight sidebands, or $9.3$~eV at the
$1.17$~eV photon energy, and the third arm at $\pi/2$ adds a four-sideband beat,
$4.7$~eV. Note that, reading the beat period off the spectrum returns the relative phase of the arms directly.

\section{Discussion}
\label{sec:discussion}

A single PINEM interaction fixes three properties of the attosecond train: when, where, and how long the peaks are. The optical phase $\arg\{g\}$ controls the arrival time, since every harmonic of the propagated density carries the
distance-independent phase $\ee^{\ii m(\arg\{g\}-\pi/2)}$ of
Eq.~\eqref{eq:harmonics_main} and all of them come into phase at
$\omega\tau_f=\arg\{g\}-\pi/2$, Eq.~\eqref{eq:tau_star_main}, independent of the value of
$|g|$. The modulus of the coupling parameter controls the remaining two. Classically the trajectories focus at $z_T/(8\pi|g|)$
\cite{baum_jap_2017,yalunin_prresearch_2021}, which holds only
while the modulation angles are small enough for the trajectories to be straight
and the caustic to be a point;
the wave treatment gives that point a finite width and moves it outward to $z_f$,
Eq.~\eqref{eq:zf_practical_main}. The same $|g|$ sets the imprinted bandwidth and therefore the duration, Eq.~\eqref{eq:fwhm_main}. A larger
coupling thus shortens the pulse and pulls the focus inward together.

At fixed excitation wavelength, coupling and phase, the electron energy enters only through
$z_T={4\pi\me v^3 \gamma^3}/\hbar\omega^2$, Eq.~\eqref{eq:zT_main}, that is through
$(v\gamma)^3$. The diffraction correction depends on $|g|$ alone and cancels
in any ratio at fixed coupling, leaving
$z_f(E_2)/z_f(E_1)=z_T(E_2)/z_T(E_1)=(\gamma_2 v_2/\gamma_1 v_1)^3$.
This ratio is what lets us compare the SEM and TEM
configurations. At
$1064$~nm this is $48.4$ for $120$~keV against $10$~keV, and $115.2$ for $200$~keV,
so at $|g|=6$ the same train forms at $63.0~\mu$m instead of $3.05$ and
$7.26$~mm, the two regimes being the bands shaded in
Fig.~\ref{fig2}(a). Nothing else changes, because neither $T$ nor
$\Delta\tau$ contains the electron energy, and therefore, the same coupling delivers the same
$101.6$~as pulse at all three energies, and the dynamics expressed in $z_T$ and
$T$ are identical.

The propagation is directly connected to the imprinted phase. The interaction leaves the sideband
amplitudes $c_N(g)$ carrying the phase $N\arg\{g\}$, and free propagation
multiplies each by $\ee^{-2\pi\ii N^2z/z_T}$, which is quadratic in $N$ and therefore
dispersive. Fourier transforming the temporal density over the retarded time $\tau$ yields the contributions of
Eq.~\eqref{eq:harmonics_main}, where the harmonic index $m$ of the optical
frequency carries the phase in the $z$-independent prefactor and the
whole dispersive part in the amplitude $J_m(4|g|\sin[2\pi mz/z_T])$. The Talbot
distance is the scale of that sine, so the $m$-th harmonic rephases with period
$z_T/m$ and the train forms where the amplitudes are jointly largest.

Multi-PINEM drives the electron with more than one interaction, and what this
adds to the single-interaction rules depends on whether the interactions act in
series or in parallel. Sequential PINEM modulates the same wave function $M$
times along its path. Specifically, for two interactions at one optical
frequency the displacement operators compose,
$\hat S(g_2)\hat S(g_1)=\hat S(g_1+g_2)$, and the pair is a single interaction
of coupling $g_1+g_2$. This is the regime of the Ramsey-type geometry of
Ref.~\citenum{echternkamp_natphys_2016}, where the two near fields sit a few
micrometres apart and the dispersion between them is negligible. Only the
dispersive phase accumulated in between, which does not commute with the
displacement, makes a sequence more than its sum. Parallel PINEM instead
divides the electron wave function among multiple arms, each characterized by
its own coupling $g_i$, so that upon coherent recombination the amplitudes add
as $\sum_i p_i c_N(g_i)$. Unless all arms are identical, this sum cannot be
represented as $c_N(g)$ for any single effective coupling $g$, so the
recombined state is inaccessible through a single PINEM interaction. The link
between duration and focal distance still holds inside each arm, but no longer
across the wave function as a whole, so $M$ arms give $M$ independent pairs
$(|g_i|,\arg\{g_i\})$ and therefore $M$ pulses, each with its own arrival time,
focal plane and duration.

Additionally, we can distinguish the nature of the multiple PINEM interactions by measuring the energy spectrum.
At equal weights, the mixture adds populations,
$P_N^{(\rm i)}=M^{-1}\sum_iJ_N^2(2|g_i|)$, and returns a Bessel comb, whereas the
which-path-erased case adds amplitudes, $P_N^{(\rm c)}\propto
M^{-1}\left|\sum_i\ee^{\ii N\arg\{g_i\}}J_N(2|g_i|)\right|^2$, a sum of squares
against the square of a sum, whose cross terms beat the envelope at a period set
by the arm phase difference. The spectrum therefore shows whether the arms were
combined as amplitudes, which is a measurement of the mutual coherence of the driving
fields.

\section{Conclusions}
\label{sec:conclusions}
This work presents photon-induced near-field interactions as a quantitative
synthesis tool for the free-electron wave function. We have studied the control
of the attosecond pulses of the electron wave function after propagation, where a
chosen position, timing, duration, and internal structure are designed on demand.
The rules come from a closed-form harmonic decomposition of the propagated
density.

The multi-PINEM platform extends the synthesis from one  set of parameters to as
many degrees of freedom as paths. Unequal couplings distribute the focus in
space, and unequal phases distribute the pulses in time. Which-path information then
decides whether the pathways recombine into a single coherent wave function, and we
identified two experimentally distinct signatures when the paths are distinguishable or not. The first is an
inter-path interference pattern that survives all the way to the Talbot revival.
The second is the spectral signature developed in either case, measured with an energy detector and independent of propagation, the telltale sign that a structured beam has been generated. 

These results open a concrete route to full wave-function engineering. The
natural extrapolation of the parallel geometry is an interaction
plane containing many independently addressed elements, namely an optical metasurface
acting on electrons, which is the electron-optical counterpart of
a spatial light modulator. Each element would contribute its own complex coupling $g_i$, which is the phase-matched Fourier component of the longitudinal near field imprinted on the individual pixels. A metasurface of such elements would therefore act as a
programmable amplitude-and-phase mask for the electron, with the multipole
coupling parameters as the design variables $|g_i|$ and $\arg\{g_i\}$. Combined with the single-interaction
rules established here, this would extend the present control to the two
transverse dimensions plus time $(2+1)$D, allowing the wave function to be shaped at will within a single compact instrument. All of this rests on the non-recoil ladder, which sets the lower end of the energy range over which such an instrument could work.

A further avenue is to drive the electron with more than one optical frequency, which adds
the repetition rate of the train a further degree of complexity, since the
temporal period is imprinted by the optical period. That
direction is already being opened from the drive side, where shaping the
instantaneous frequency and envelope of the light is predicted to give
near-arbitrary control of the sideband structure in time
\cite{lastovickovastreshkova_prresearch_2026}; combined with the per-path
addressing described here, the drive and the geometry would become independent
design parameters. Together with the
coherence metrology discussed here, these
extensions point toward electron microscopes in which the probe itself is a
synthesized quantum object, designed for quantum optics, sensing, or future ultra-fast metrology applications.

\section*{Methods}
\label{sec:methods}

\subsection*{Interaction Hamiltonian and the PINEM scattering operator}

Following the formalism introduced in Ref.~\cite{feist_nature_2015}, we describe an electron of charge $-e$ interacting with a localized,
time-dependent optical near field through the minimal-coupling Hamiltonian
$\hat H(t)=[\hat{\mathbf p}+e\mathbf A(\mathbf r,t)]^2/2m-e\varphi(\mathbf r,t)$,
in the Coulomb gauge and with the scalar potential $\varphi=0$ in the source-free
interaction region.
Expanding the kinetic term separates the linear coupling
\begin{equation}
\hat{H}_{\mathrm{int}}(t)=\frac{e}{2m}\Big(\hat{\mathbf p}\cdot\mathbf A+\mathbf A\cdot\hat{\mathbf p}\Big)
\label{eq:Hint_linearA}
\end{equation}
from the ponderomotive term $e^2\mathbf A^2/2m$. The sideband ladder
is generated by Eq.~\eqref{eq:Hint_linearA}; $\mathbf A^2$ produces an elastic
phase shift and a weak $2\omega$ modulation, negligible for field amplitudes below
$E_{\mathrm{thres}}=2m\gamma v\omega/e\sim10^{12}$~V/m, which is orders of
magnitude above typical experimental fields
\cite{digiulio_optica_2019,garciadeabajo_prl_2021}. For an
electron travelling along $z$ the coupling reduces
to its longitudinal component, $\hat H_{\rm int}=(e/2m)[\hat p_z A(z,t)+A(z,t)\hat p_z]$
with $A(z,t)=E_z(z)\sin(\omega t)/\omega$.

We work in the energy-ladder basis $\{|N\rangle\}$ with
$E_N=E_0+N\hbar\omega$ and ladder operators $\hat a^\dagger|N\rangle=|N+1\rangle$,
$\hat a|N\rangle=|N-1\rangle$. In the equal-Rabi approximation appropriate to a
prescribed classical field, all adjacent transitions share the same coupling and
$[\hat a,\hat a^\dagger]=0$. Evaluating Eq.~\eqref{eq:Hint_linearA} between
plane-wave ladder states and using $E_0\gg|N|\hbar\omega$ to replace
$k_{N+1}+k_N\simeq2k_0$ and $\Delta k_N\simeq\omega/v$ gives an $N$-independent
nearest-neighbour matrix element controlled by the phase-matched Fourier component
of the longitudinal field,
\begin{equation}
g(\omega)=\frac{e}{2\hbar\omega}\int_{-\infty}^{\infty}dz\,E_z(z,\omega)\,\ee^{-\ii(\omega/v)z}.
\label{eq:gomega_def}
\end{equation}
Because $[\hat a,\hat a^\dagger]=0$ the time ordering can be dropped from the scattering matrix and the
$\pm2\omega$ terms average out over a cycle, leaving the displacement operator
\begin{equation}
\hat S=\exp\!\left(g\,\hat a^\dagger-g^*\,\hat a\right),
\label{eq:S_displacement}
\end{equation}
whose action on $|0\rangle$ gives Eq.~\eqref{eq:cN_main}
\cite{garciadeabajo_nanolett_2010,park_njp_2010,
feist_nature_2015}. The derivation of the Bessel amplitudes from
Eq.~\eqref{eq:S_displacement} is given in Sec.~S1 of the SI, Eqs.~(S1)--(S4). Higher sidebands
are not driven by direct $\Delta N>1$ matrix elements but build up from repeated
single-photon steps within one interaction [Eq.~(S2) of the SI]. The overall sign of $g$ is convention
dependent; all results below depend only on $|g|$ and on phase \emph{differences}
between couplings. The nonrecoil ladder assumed here is well justified at the
energies considered: quantum corrections beyond it scale with $\hbar\omega/E_0$,
which is $1.2\times10^{-4}$ at $10$~keV and $1064$~nm, and become significant only
for electron energies in the eV to hundred-eV range
\cite{synanidis_sciadv_2024}.

\subsection*{Dispersion, the Talbot distance, and free propagation}

Representing $|N\rangle$ as a longitudinal plane wave with wave number
$k_N=\sqrt{(E_N+mc^2)^2-(mc^2)^2}/\hbar c$ and expanding about $N=0$ using
$\partial k/\partial E|_{E_0}=1/\hbar v$ and
$\partial^2k/\partial E^2|_{E_0}=-1/\hbar\gamma^3mv^3$ gives
\begin{equation}
k_N\simeq k_0+\frac{N\omega}{v}-\frac{2\pi N^2}{z_T}+\cdots,
\label{eq:kN_talbot_form}
\end{equation}
with the Talbot distance $z_T$ defined in Eq.~\eqref{eq:zT_main}. The linear term is a uniform group
delay, removed by the retarded time $\tau=t-z/v$; the quadratic term is the
dispersion that drives the reshaping. Factoring out the carrier phase,
\begin{equation}
\psi(z,\tau)\propto\sum_{N}c_N(g)\,\ee^{-\ii N\omega\tau}\,\ee^{- 2\pi \ii N^2z/z_T},
\label{eq:psi_spacetime}
\end{equation}
which is the working prescription used throughout [Eqs.~(S5) and (S17) of the SI]: the coefficients are fixed at
the interaction and propagation multiplies each order by a quadratic phase.

\subsection*{Harmonic decomposition and focus position}

Inserting Eq.~\eqref{eq:cN_main} into Eq.~\eqref{eq:psi_spacetime} and applying
Graf's addition theorem [Eq.~(9.1.79) in Ref.~\citenum{AS1972}] to
$n(z,\tau)=|\psi|^2$ gives Eq.~\eqref{eq:harmonics_main}, the steps
being Eqs.~(S5)--(S11) of Sec.~S2 in the SI. Pairing the terms $\pm m$, Eq.~(S12), gives the equivalent real form
\begin{align}
n(z,\tau)
=\;&
1+2\sum_{m=1}^{\infty}
J_m\!\left(4|g|\sin\frac{2\pi m z}{z_T}\right)
\nonumber\\
&\times
\cos\!\left[m\left(\omega\tau-\phi+\frac{\pi}{2}\right)\right].
\label{eq:carpet_harmonics_real}
\end{align}
We verified Eq.~\eqref{eq:harmonics_main} against a discrete Fourier transform of
the numerically propagated density: the two agree to $3\times10^{-14}$ [see Sec.~S6 of the SI].

The propagated wave function $\psi(z,\tau)$ can equivalently be written as the
diffraction integral of,
\begin{equation}
\psi \propto
\int\frac{d\tau'}{\sqrt{\ii u}}\;
\exp\!\left\{2\ii|g|\left[
\frac{\omega^{2}(\tau-\tau')^{2}}{2u}-\sin(\omega\tau'-\phi)
\right]\right\} ,
\label{eq:kirchhoff_main}
\end{equation}
with $u=8\pi|g|z/z_T$ [Eqs.~(S15)-(S20) of the SI] the propagation distance in units of the classical
compression distance. This is Kirchhoff diffraction from a sinusoidal phase
grating of modulation depth $2|g|$, the retarded time $\tau'$ at which the
electron crossed the interaction region playing the role of the coordinate across
the grating. Integrals of this form define the Pearcey function of cusp
diffraction, and it is from it that both the focus position of
Eq.~\eqref{eq:zf_practical_main} and the pulse duration of
Eq.~\eqref{eq:fwhm_main} follow [Eqs.~(S21)--(S25) of Sec.~S3 and Eqs.~(S29)--(S35) of Sec.~S4 of the SI].

\subsection*{Parallel paths}

The parallel-path results follow from Eq.~\eqref{eq:rho_after_main}. Densities
are computed as $\rho=|\sum_i p_i\,\psi_i|^2$ in the which-path-erased case and
$\rho=\sum_i|p_i|^2|\psi_i|^2$ in the statistical mixture, with equal weights
$|p_i|^2=1/M$ throughout, each normalized to its own incident-plane mean so that
${\rm Tr}\,\{\rho\}=1$; for the coherent case this normalization is exactly
$\mathcal N_\coh^2$, with the closed form [see Eq.~(S48)]
\begin{equation}
\mathcal N_\coh^{-2}
=\sum_{i,j}\Gamma_{ij}\,\langle0|\hat S_j^\dagger\hat S_i|0\rangle
=\sum_{i,j}\Gamma_{ij}\,J_0\!\left(2|g_i-g_j|\right),
\label{eq:Ncoh_methods}
\end{equation}
which follows from the composition property
$\hat S_j^\dagger\hat S_i=\hat S(g_i-g_j)$ of the displacement operators; for
the two-arm example of Fig.~\ref{fig4} it evaluates to
$\mathcal N_\coh^{-2}=1+J_0(2|g_1-g_2|)=1.0093$. The sideband
populations of the two cases, Eqs.~(S44) and (S46) of the SI,
\begin{align}
P_N^{(\rm i)}=&\sum_i |p_i|^2 \left| c_N(g_i)\right|^2,
\\
P_N^{(\rm c)}=&\mathcal N_\coh^2\left|\sum_i p_i\,c_N(g_i)\right|^2 ,
\label{eq:PN_paths_methods}
\end{align}
differ in how the arms combine at the level of the amplitudes $c_N(g_i)$, so that the relative optical phases
$\arg\{g_i\}$ enter the spectrum only in the second. Section~S5 of the
SI, Eqs.~(S41)--(S47), reduces Eq.~\eqref{eq:rho_after_main} to both limits.
Additionally, the intermediate case of
partial path coherence follows from scaling the off-diagonal elements of
$\Gamma$ by a factor $\mu\in[0,1]$, which interpolates linearly between them
while keeping $\Gamma$ a valid density matrix.


Energy spectra are obtained by broadening the discrete populations with a Lorentzian of
$2\delta = 0.7$~eV spectral full-width at half maximum, such that $I(E)=\sum_N P^{(\rm x)}_N\,{(\delta/\pi)}/{[(E-N\hbar\omega)^2+\delta^2]}$.

\subsection*{Harmonic decomposition of parallel paths}

The spectrum and the charge density of the recombined electron follow from the density matrix
Eq.~\eqref{eq:rho_after_main}. Propagation imprints the quadratic phase of
Eq.~\eqref{eq:psi_spacetime} to each of its two indices, so that
\begin{equation}
\rho_{NN'}^{(\rm x)}(z)=\rho_{NN'}^{(\rm x)}(0)\,
\ee^{-2\pi\ii(N^{2}-N'^{2})z/z_T} ,
\label{eq:rho_prop_methods}
\end{equation}
where ${\rm x}={\rm i}$ stands for the statistical mixture of
Eq.~\eqref{eq:rho_mixture_main} and ${\rm x}={\rm c}$ for the which-path-erased
state of Eq.~\eqref{eq:rho_coherent_main}. The temporal
structure lies in the coherences between sidebands separated by $m=N-N'$ photon
energies, each of them oscillating at $m\omega$, and expanding the charge density in harmonics of the optical period, we have that
\begin{align}
n^{(\rm x)}(z,\tau)
&=\sum_m \mathcal I_m^{(\rm x)}(z)\,\ee^{-\ii m\omega\tau},
\nonumber\\
\mathcal I_m^{(\rm x)}(z)
&=\sum_N\rho_{N+m,N}^{(\rm x)}(z) .
\label{eq:n_paths_methods}
\end{align}
From the general $\rho_{NN'}^{(\rm x)}$ of
Eq.~\eqref{eq:rho_after_main} and the coefficients of Eq.~\eqref{eq:cN_main}, with
$\phi_i=\arg\{g_i\}$, we obtain one term for every ordered pair of arms $(i,j)$,
\begin{align}
\mathcal I_m^{(ij)}(z)
=\;&\ee^{\ii m\phi_i}\,\ee^{-2\pi\ii m^{2}z/z_T}
\nonumber\\
&\times\sum_N J_{N+m}\!\left(2|g_i|\right)
J_N\!\left(2|g_j|\right)\ee^{\ii N\theta_{ij}} ,
\label{eq:Im_pair_methods}
\end{align}
in which the propagation distance appears twice, once in the
overall phase and once in the relative angle
$\theta_{ij}=\phi_i-\phi_j-4\pi m z/z_T$ that weights the sideband sum. The two
recombinations differ only in the weight given to each pair,
\begin{align}
\mathcal I_m^{(\rm i)} (z)
&=\sum_i |p_i|^{2}\,\mathcal I_m^{(ii)},\\
\mathcal I_m^{(\rm c)} (z)
&=\mathcal N_\coh^{2}\sum_{i,j} p_ip_j^*\,
\mathcal I_m^{(ij)} ,
\label{eq:Im_paths_methods}
\end{align}
the weights being the elements of the path density matrix, which a
marked path reduces to its diagonal, $\Gamma_{ij}=|p_i|^{2}\delta_{ij}$, so that
only the terms with $i=j$ survive, whereas a fully coherent path state, $\Gamma_{ij}=p_ip_j^*$, keeps all of them. That which-path erased form factorizes
the double sum over arms, so that the coherent harmonics contract the recombined
amplitudes directly,
\begin{align}
\mathcal I_m^{(\rm c)}(z)
&=\mathcal N_\coh^{2}\sum_N \Psi_{N+m}(z)\,\Psi_N^{*}(z),\\
\Psi_N(z)
&=\sum_i p_i\,c_N(g_i)\,\ee^{-2\pi\ii N^{2}z/z_T} ,
\label{eq:Im_coh_amp_methods}
\end{align}
where $\Psi_N(z)$ is the propagated sideband amplitude of the
recombined wave function.
Partial path coherence scales the off-diagonal elements of $\Gamma$ by $\mu\in [0,1]$ and its propagation follows from the general weights of Eq.~\eqref{eq:Im_paths_methods} alone.

Carrying out the sideband sum of Eq.~\eqref{eq:Im_pair_methods}
with Graf's addition theorem reduces the difference between
the two recombinations to a single phase. Within one arm the angle $\theta_{ii}$
enters through a combination that cancels the quadratic phase of propagation, so
that every diagonal term returns the single-arm harmonic of
Eq.~\eqref{eq:harmonics_main} and the mixture is built from carpets that hold
their position in retarded time while their harmonic amplitudes vary along $z$.
Unequal couplings break that cancellation, their cross terms retaining a phase
that advances with distance, so that the interference pattern of the erased case
moves in retarded time as the electron travels. We give the details in Sec.~S5~A of the SI.

\subsection*{Numerics}

All figures are computed from Eqs.~\eqref{eq:cN_main} and
\eqref{eq:psi_spacetime} with sideband cutoffs $|N|\leq110$, sufficient for the
couplings used here ($\sum_N P_N=1$ to $10^{-12}$). Focus positions quoted as
``numerical'' are the argmax of the propagated density over a fine longitudinal
grid; those quoted as analytic follow from Eq.~\eqref{eq:zf_practical_main}. The
two agree to $0.01\%$ at $|g|=6$. Pulse durations are full widths at half maximum of
the individual peaks of the focal cut. Densities are normalized to the incident
mean, which equals the trace and is conserved along $z$ to machine precision in all
calculations reported here. The implementation, together with the identity and
convergence tests used to validate it, is described in Sec.~S6 of the SI.

\begin{acknowledgments}
The authors thank F.~Javier Garc\'{i}a de Abajo, Claus Ropers,
and Sophie Meuret for insightful discussions.

This work is part of the research program of the Dutch Research Council NWO. It has received funding from the European Research Council (ERC) under the European Union’s Grant Agreement No. 101019932 Quantum Electron Wavepacket Spectroscopy (QEWS) and Grant Agreement No. 101017720 (EBEAM). 

\end{acknowledgments}

\bibliographystyle{apsrev4-2}
\bibliography{refs_paper1}

\end{document}


\title{Supplementary Information for\\
``Synthesizing free-electron wave functions by stimulated near-field interactions''}

\author{\'{A}lvaro~Rodr\'{i}guez~Echarri}
\affiliation{Center for Nanophotonics, NWO Institute AMOLF, 1098 XG Amsterdam, The Netherlands}
\author{Albert~Polman}
\affiliation{Center for Nanophotonics, NWO Institute AMOLF, 1098 XG Amsterdam, The Netherlands}

\date{\today}
\maketitle

\tableofcontents

\section{Bessel-law sideband amplitudes}
\label{si:bessel}

We derive Eq.~(1) of the main text from the displacement-operator form of the
scattering matrix, following the standard PINEM treatment
\cite{garciadeabajo_nanolett_2010,park_njp_2010}. In particular, following the Ref.~\cite{feist_nature_2015} formalism, in the equal-Rabi approximation, the ladder operators commute $[\hat a,\hat a^\dagger]=0$, so that
\begin{equation}
\hat S=\exp\!\left(g\hat a^\dagger-g^*\hat a\right)
=\exp\!\left(g\hat a^\dagger\right)\exp\!\left(-g^*\hat a\right),
\label{eq:si_factorize}
\end{equation}
given by the Baker--Campbell--Hausdorff formula.
Expanding both exponentials and acting on $|0\rangle$,
\begin{align}
\hat S|0\rangle
&=
\sum_{r=0}^{\infty}\sum_{s=0}^{\infty}
\frac{g^{r}}{r!}\frac{(-g^*)^{s}}{s!}\,
(\hat a^\dagger)^{r}\hat a^{s}|0\rangle
\nonumber\\
&=
\sum_{r,s}
\frac{g^{r}(-g^*)^{s}}{r!\,s!}\,|r-s\rangle .
\label{eq:si_double_sum}
\end{align}
Collecting the terms with $r-s=N$ and writing $g=|g|\ee^{\ii\phi}$,
\begin{equation}
c_N=\langle N|\hat S|0\rangle
=
\ee^{\ii N\phi}
\sum_{s=0}^{\infty}
\frac{(-1)^{s}\,|g|^{\,N+2s}}{(N+s)!\,s!},
\label{eq:si_cN_series}
\end{equation}
which is the series expansion of the Bessel function of the first kind
[Eq.~(9.1.10) in Ref.~\citenum{AS1972}] evaluated at argument $2|g|$,
$J_N(x)=\sum_s(-1)^s(x/2)^{N+2s}/[s!(N+s)!]$. Hence
\begin{equation}
c_N(g)=\ee^{\ii N\phi}J_N\!\left(2|g|\right)
=\left(\frac{g}{|g|}\right)^{\!N}J_N\!\left(2|g|\right),
\label{eq:si_cN}
\end{equation}
and $P_N=J_N^2(2|g|)$. Additionally, conservation of population is guaranteed by the Bessel sum rule $\sum_N J_N^2(x)=1$ [Eq.~(9.1.76) in Ref.~\citenum{AS1972}].

\section{Harmonic decomposition of the propagated density}
\label{si:harmonics}

This note derives Eqs.~(4) and (19) of the main text. Free propagation multiplies each sideband by the quadratic phase of Eq.~(18) of the main text, so that the density in the co-moving frame reads
\begin{align}
n(z,\tau)=|\psi|^2
=\sum_{N,N'}c_N^{\phantom{*}}c^*_{N'}\,
&\ee^{-\ii(N-N')\omega\tau}\ee^{-2\pi\ii(N^2-N'^2)z/z_T}.
\label{eq:si_double_ladder}
\end{align}
Only the difference of the two ladder indices enters the retarded time. Because
that difference is an integer, $n$ is periodic in $\tau$ with the optical period
$T=2\pi/\omega$ and is fixed by its Fourier coefficients,
\begin{equation}
n(z,\tau)=\sum_{m=-\infty}^{\infty}\mathcal I_m(z)\,\ee^{-\ii m\omega\tau},
\qquad
\mathcal I_m(z)=\frac{1}{T}\int_{-T/2}^{T/2}\!\!d\tau\;n(z,\tau)\,
\ee^{\ii m\omega\tau}.
\label{eq:si_fourier_pair}
\end{equation}
Projecting Eq.~\eqref{eq:si_double_ladder} with the second of these keeps only
the terms with $N-N'=m$, so we label them by the harmonic order $m$ and sum over
the remaining index $N'$. First, we note that the coupling phases $c_N=\ee^{\ii N\phi}J_N(2|g|)$ from
Eq.~\eqref{eq:si_cN} collapse,
\begin{equation}
c^{\phantom{*}}_{N'+m}c^*_{N'}=\ee^{\ii m\phi}\,J_{N'+m}(2|g|)\,J_{N'}(2|g|).
\label{eq:si_phase_collapse}
\end{equation}
Second, the quadratic propagation phase factorizes,
\begin{equation}
N^2-N'^2=(N-N')(N+N')=m\,(2N'+m),
\label{eq:si_quadratic_split}
\end{equation}
which separates a contribution that is constant in the summation index from one
that is linear in it. After renaming $N'\to N$, writing $\theta=-4\pi mz/z_T$ and inserting
Eqs.~\eqref{eq:si_phase_collapse} and \eqref{eq:si_quadratic_split} into
Eq.~\eqref{eq:si_double_ladder},
\begin{equation}
\mathcal I_m(z)=\ee^{\ii m\phi}\,\ee^{-2\pi\ii m^2z/z_T}
\sum_{N}J_{N+m}(2|g|)\,J_N(2|g|)\,\ee^{\ii N\theta},
\label{eq:si_Im_presum}
\end{equation}
The exponent left
inside the sum is linear in $N$, which is precisely the structure that Graf's
addition theorem [Eq.~(9.1.79) in Ref.~\citenum{AS1972}], which applies as
\begin{align}
\sum_{N}J_{N+m}(2|g|)&\,J_N(2|g|)\,\ee^{\ii N\theta}
=\ee^{\ii m(\pi-\theta)/2}\,
J_m\!\left(4|g|\sin\frac{\theta}{2}\right).
\label{eq:si_graf_applied}
\end{align}
Substituting Eq.~\eqref{eq:si_graf_applied} into Eq.~\eqref{eq:si_Im_presum} with
$\theta=-4\pi mz/z_T$, the Graf prefactor splits as
$\ee^{\ii m(\pi-\theta)/2}=\ee^{\ii m\pi/2}\,\ee^{2\pi\ii m^2z/z_T}$, whose second
factor cancels the $\ee^{-2\pi\ii m^2 z/z_T}$ standing outside the sum. In the Bessel argument
$\sin(\theta/2)=-\sin(2\pi mz/z_T)$, and $J_m(-y)=(-1)^mJ_m(y)$ turns that sign
into a further factor $\ee^{\ii m\pi}$. The three surviving phases combine as $\ee^{\ii m\phi}\,\ee^{\ii m\pi/2}\,\ee^{\ii m\pi}
=\ee^{\ii m(\phi+3\pi/2)}=\ee^{\ii m(\phi-\pi/2)}$
for integer $m$, resulting in 
\begin{equation}
\mathcal I_m(z)=\ee^{\ii m(\phi-\pi/2)}\,
J_m\!\left(4|g|\sin\frac{2\pi mz}{z_T}\right),
\label{eq:si_Im}
\end{equation}
which is Eq.~(4) of the main text. Since $J_{-m}(-y)=J_m(y)$ for integer $m$, Eq.~\eqref{eq:si_Im} satisfies
$\mathcal I_{-m}=\mathcal I_m^*$, as a real density requires, and
$\mathcal I_0=J_0(0)=1$. Pairing $\pm m$ in
$n=\sum_m\mathcal I_m\ee^{-\ii m\omega\tau}$ then gives
\begin{align}
n(z,\tau)
=\;&1+2\sum_{m=1}^{\infty}
J_m\!\left(4|g|\sin\frac{2\pi mz}{z_T}\right)
\cos\!\left[m\left(\omega\tau-\phi+\frac{\pi}{2}\right)\right],
\label{eq:si_carpet}
\end{align}
which is Eq.~(19) of the main text. The $m=0$ term carries the normalization, so
$\langle n\rangle_\tau=1$ in every plane, as the conservation of the trace under
free propagation requires. Two limits follow at once. The density is flat at
$z=0$, where every Bessel argument vanishes (i.e., $J_m(0)=0$ for $m>0$), confirming that the interaction
imprints pure phase and leaves the density untouched; and it is flat again
wherever $\sin(2\pi mz/z_T)$ vanishes for all $m$ simultaneously, that is at
$z=z_T/2$ and $z=z_T$.

Note that every density written here is divided by the mean of the incident plane,
$\langle n\rangle_\tau=1$, a value that free propagation conserves, so that $n$ is a
dimensionless profile along the retarded time rather than a charge per unit length.
Absolute units are recovered by multiplying it by $-e$ (i.e., the electron charge) and by the incident electron
density.

\section{Diffraction analysis of the temporal focus}
\label{si:pearcey}

This note derives Eq.~(6) of the main text, continuing from the harmonic
decomposition of Sec.~\ref{si:harmonics}.
We want to derive an expression for the peak position of the charge density at the focus, which might depend on the coupling, the electron energy, or the drive wavelength. 
The direct route, setting $\partial n/\partial z=0$, does not work because the
resulting series has no closed-form. Incidentally, linearization of the sinusoidal modulation removes the maximum altogether and leaves only the classical result, as we shall see below.

\begin{figure}[b]
\centering
\includegraphics[width=\linewidth]{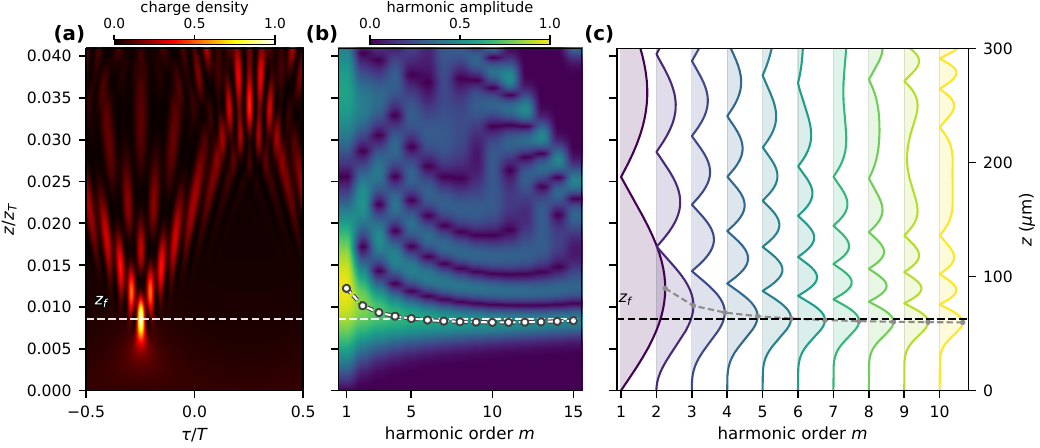}
\caption{\textbf{Decomposition of the attosecond pulse.} The maps of
Fig.~2(b,c) of the main text over the first $300~\mu$m, that is $0.041\,z_T$ or
$4.8\,z_f$, for $|g|=6$, $10$~keV electrons and $1064$~nm driving wavelength.
The three panels share the propagation axis, given as $z/z_T$ on the left and in
micrometers on the right. \textbf{(a)} Charge density against retarded time.
\textbf{(b)} Harmonic amplitudes $|\mathcal I_m|$ of Eq.~\eqref{eq:si_Im} for
$m=1$ to $15$. \textbf{(c)} The same amplitudes for $m=1$ to $10$ as curves,
each offset by its order and drawn to a common scale. The line with circles joins
the first maximum of every order, and the dashed line is $z_f$ obtained from the maximum numerically.}
\label{figS_zoom}
\end{figure}

\begin{figure}[b]
\centering
\includegraphics[width=0.52\linewidth]{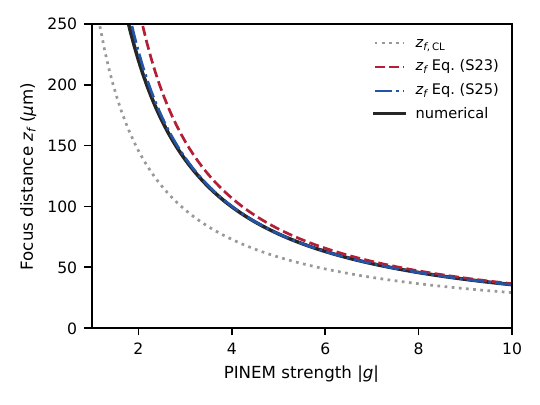}
\caption{\textbf{Focus distance against coupling strength.}
Classical compression distance $z_{f,\rm CL}$ of Eq.~\eqref{eq:si_zcl} (dotted gray),
the derived result of Eq.~\eqref{eq:si_zf_pearcey} (dashed red), the fitted
formula of Eq.~\eqref{eq:si_zf_practical} (dash-dotted blue) and direct maximization
of the propagated density $n(z,\tau)$ at $\tau_f$ (solid black), for $10$~keV electrons and
$1064$~nm driving wavelength.}
\label{figS_zfcheck}
\end{figure}

\paragraph{Maximization of the charge density.}
The phases in
Eq.~\eqref{eq:si_carpet} are linear in $m$ with a slope set by $\phi$ alone, so
every cosine reaches unity at the same retarded times,
\begin{equation}
\omega\tau_f=\phi-\frac{\pi}{2}+2\pi\ell,
\qquad\ell=0,\pm1,\pm2,\dots,
\label{eq:si_inphase}
\end{equation}
which is Eq.~(5) of the main text, and it does not depend on $z$. At this point,
Eq.~\eqref{eq:si_carpet} reads
\begin{equation}
n(z,\tau_f)=1+2\sum_{m=1}^{\infty}J_m\!\left(4|g|\sin\frac{2\pi mz}{z_T}\right).
\label{eq:si_focus_exact}
\end{equation}
The attosecond peak
forms where the sum is largest. In the pre-revival region $z\ll z_T$ the sine
linearizes, $4|g|\sin(2\pi mz/z_T)\simeq m\,u$, yielding
\begin{equation}
n(z,\tau_f)=1+2\sum_{m=1}^{\infty}J_m\!\left(m\,u\right),
\qquad \text{with} \qquad
u=\frac{8\pi|g|z}{z_T}.
\label{eq:si_klystron}
\end{equation}
Equation~\eqref{eq:si_klystron} is the classical bunching form of a
velocity-modulated beam undergoing ballistic compression
\cite{kutsaev2021electron,pan2023weak}. Here, $u$ is the common scaled
propagation coordinate, while the amplitude of harmonic $m$ is $J_m(mu)$.
Because the first maximum of $J_m(y)$ lies at
$j'_{m,1}=m+0.8086\,m^{1/3}+\mathcal O(m^{-1/3})$ [Eq.~(9.5.16) in
Ref.~\citenum{AS1972}], the harmonics peak at
$u_m=j'_{m,1}/m=1+0.8086\,m^{-2/3}+\mathcal O(m^{-4/3})$ and converge towards
the classical compression point $u=1$. We therefore denote the corresponding
distance by
\begin{equation}
z_{f,\mathrm{CL}}\equiv\frac{z_T}{8\pi|g|},
\label{eq:si_zcl}
\end{equation}
in agreement with Refs.~\citenum{baum_jap_2017} and
\citenum{yalunin_prresearch_2021}. Figure~\ref{figS_zoom} resolves the
finite-order spread by marking the first maximum of each harmonic on the map
and on the individual curves. However, Fig.~\ref{figS_zfcheck} shows that, at $|g|=6$,
$z_{f,\mathrm{CL}}=48.6~\mu\mathrm{m}$, whereas direct numerical
maximization gives $z_f=63.0~\mu\mathrm{m}$. Thus, the classical estimate
falls approximately $23\%$ short of the numerical focus. It remains shorter
over the full range of couplings considered, with the relative discrepancy
decreasing only slowly as $|g|$ increases. The maximization of the charge density is therefore not accurate enough, and we approach the problem differently below, returning to the wave function rather than to the density. As we have seen above, in this case, the problem is that harmonic $m$ evolves on its own length scale, and a first-order general expression does not work.

\paragraph{Maximum of the wave function.}
We start from the wavefunction of the electron in Eq.~(18) of the main text,
\begin{equation}
\psi(z,\tau)\propto\sum_{N} J_N(2|g|)\,\ee^{\ii N\phi}\,\ee^{-\ii N\omega\tau}\,
\ee^{-\ii\gamma N^{2}},
\qquad
\gamma\equiv\frac{2\pi z}{z_T}=\frac{u}{4|g|},
\label{eq:si_psi_prop}
\end{equation}
whose propagation factor is quadratic in the summation index. This is what
prevents the sum from being carried out, since the Bessel generating functions
sum only phases linear in $N$, and therefore we have to apply some tricks before using the corresponding Jacobi--Anger expression. A quadratic phase can nevertheless be traded for a linear one at the cost of one integration: completing the square in
$s^{2}/4\gamma-Ns=(s-2\gamma N)^{2}/4\gamma-\gamma N^{2}$ and using
$\int dY\,\ee^{\ii Y^{2}/4\gamma}=\sqrt{4\pi\ii\gamma}$,
\begin{equation}
\ee^{-\ii\gamma N^{2}}
=\frac{1}{\sqrt{4\pi\ii\gamma}}\int_{-\infty}^{\infty} dx\;\ee^{\,\ii x^{2}/4\gamma}\;\ee^{-\ii Nx},
\label{eq:si_gauss_id}
\end{equation}
which is exact for every $N$, the price of removing $N^{2}$ from the exponent
being the auxiliary variable $x$. Inserting Eq.~\eqref{eq:si_gauss_id} into
Eq.~\eqref{eq:si_psi_prop} and exchanging the sum with the integral leaves an
exponent linear in $N$, so that the sum is the Jacobi--Anger expansion evaluated
at the displaced time $\tau+x/\omega$,
\begin{equation}
\sum_{N}J_N(2|g|)\,\ee^{-\ii N(\omega\tau+x-\phi)}
=\ee^{-2\ii|g|\sin(\omega\tau+x-\phi)}
=\psi_0\!\left(\tau+\frac{x}{\omega}\right),
\label{eq:si_jacobi}
\end{equation}
with $\psi_0(\tau)=\ee^{-2\ii|g|\sin(\omega\tau-\phi)}$ the wave function at
$z=0$, of unit modulus. The summation has disappeared, and substituting
$x=\omega(\tau'-\tau)$, leaves
\begin{equation}
\psi(z,\tau)\propto
\frac{1}{\sqrt{\ii u}}
\int d\tau'\;
\exp\!\left\{2\ii|g|\left[
\frac{\omega^{2}(\tau-\tau')^{2}}{2u}-\sin(\omega\tau'-\phi)
\right]\right\} .
\label{eq:si_kirchhoff}
\end{equation}
This is Kirchhoff diffraction from a sinusoidal phase grating of modulation depth
$2|g|$, the optical form of the longitudinal Talbot problem. The auxiliary
variable $\tau'$ has become the time at which the electron crossed the interaction
region, and the coupling enters only as the large parameter $2|g|$ multiplying the
exponent.
We now evaluate Eq.~\eqref{eq:si_kirchhoff} at the in-phase time
$\omega\tau_f=\phi-\pi/2$ of Eq.~\eqref{eq:si_inphase} and use the rescaled variable
\begin{equation}
s=\left(\frac{|g|}{12}\right)^{1/4}\omega(\tau'-\tau_f),
\label{eq:si_svar}
\end{equation}
so that $\sin(\omega\tau'-\phi)=-\cos[(12/|g|)^{1/4}s]$ and the exponent of
Eq.~\eqref{eq:si_kirchhoff} is even in $s$. Expanding the cosine through its
quartic term, and normalizing the expression to $s^4$, we obtain $ \psi(z,\tau_f)\propto
 \ee^{2\ii|g|}\,\mathcal P(B)/{\sqrt{\ii u}}$ the on-axis Pearcey function
\begin{equation}
\mathcal P(B)
=
\int_{-\infty}^{\infty} ds\,
\ee^{\ii\left(Bs^{2}+s^{4}\right)},
\qquad \text{with} \qquad
B=\sqrt{6}\,(2|g|)^{1/2}\,\frac{1-u}{u},
\label{eq:si_pearcey_def}
\end{equation}
in which the coupling strength, electron energy and drive frequency have all been
absorbed into the single scaled detuning $B$.
The on-axis density keeps the factor of Eq.~\eqref{eq:si_kirchhoff},
$n(z,\tau_f)\propto |\mathcal P(B)|^{2}/(\pi u)$, in which $1/u$ is a slowly
varying envelope next to the cusp, so locating the focus becomes a one-parameter
optimization, for which the solution is known to be $B^* = -2.199$, the envelope
displacing it only at the next order that we shall account for later.
Solving for $B=B^*$ in Eq.~\eqref{eq:si_pearcey_def} gives $u$ in closed
form, and with $z=u\,z_{f,\rm CL}$, we have that 
\begin{equation}
z_f
=
z_{f,\rm CL}
\left[1-\frac{0.898}{\sqrt{2|g|}}\right]^{-1},
\qquad
\frac{|B^*|}{\sqrt{6}}=0.898,
\label{eq:si_zf_pearcey}
\end{equation}
so the classical estimate is approached only as $|g|\to\infty$, and with a
characteristic diffractive power $|g|^{-1/2}$.

However, Eq.~\eqref{eq:si_zf_pearcey} does not reproduce the numerical focus. As Fig.~\ref{figS_zfcheck} shows, it overestimates it by $7.1\%$ at $|g|=4$ and by $4.2\%$ at $|g|=6$,
the discrepancy starting at order $(2|g|)^{-1}$, since the leading coefficient
is the correct one. The reason is
contained in Eq.~\eqref{eq:si_pearcey_def}: the cosine was truncated after its
quartic term, and what was discarded there, together with the slowly varying
prefactor $1/u$, contributes to $z_f$ at the
same order as the second term of the expansion of Eq.~\eqref{eq:si_zf_pearcey}, which expanded reads as
\begin{equation}
z_f=z_{f,\rm CL}\left[1+\frac{0.898}{\sqrt{2|g|}}
+\frac{0.898^{2}}{2|g|}+\dots\right].
\label{eq:si_zf_expanded}
\end{equation}
Keeping the derived leading term and absorbing those two contributions into
the second coefficient, which is determined from the exact maxima of the
harmonic sum, gives the practical formula quoted in the main text,
\begin{equation}
z_f
=
z_{f,\rm CL}
\left[
1+\frac{0.898}{\sqrt{2|g|}}+\frac{0.44}{2|g|} + \mathcal O\big((2|g|)^{-3/2}\big)
\right],
\label{eq:si_zf_practical}
\end{equation}
whose largest relative error over that window is $0.38\%$, at $|g|=4$, and
better than $0.14\%$ for $|g|\geq6$.

\begin{figure}[b]
\centering
\includegraphics[width=\linewidth]{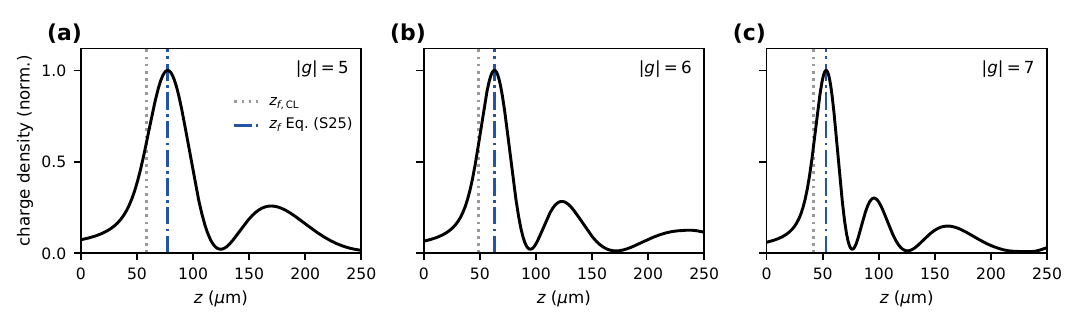}
\caption{\textbf{Formation of the attosecond peak along the
focus.} Charge density at the in-phase time $\tau_f=-T/4$, that is
Eq.~\eqref{eq:si_inphase} with $\phi=0$, as a function of propagation distance
for \textbf{(a)} $|g|=5$, \textbf{(b)} $|g|=6$ and \textbf{(c)} $|g|=7$, all at
$10$~keV electrons and $1064$~nm driving wavelength. Each curve is normalized to its own maximum. The dotted line marks the classical compression distance
$z_{f,\rm CL} = z_T/(8\pi|g|)$ with $z_T = 7.33$~mm, and the dashed line the diffraction-corrected $z_f$ of
Eq.~\eqref{eq:si_zf_practical}.}
\label{figS_focus}
\end{figure}

For the $120$~keV, $800$~nm combination at $|g|=6$ the classical estimate gives
$1.33$~mm, Eq.~\eqref{eq:si_zf_pearcey} gives $1.80$~mm and
Eq.~\eqref{eq:si_zf_practical} gives $1.72$~mm, the latter coinciding with both the
exact maximum of the harmonic sum and full numerical propagation. For the
$10$~keV, $1064$~nm system of this work we obtain $48.6~\mu$m,
$65.6$~$\mu$m and $63.0~\mu$m, respectively.
Figure~\ref{figS_focus} follows the density at $\tau_f$ along the
focus and shows the correction. The maxima fall at $77.4$, $63.0$ and
$53.0~\mu$m for $|g|=5$, $6$ and $7$, within $0.1\%$  of
Eq.~\eqref{eq:si_zf_practical} in every case, while the classical estimate lies  $20\%$ below the actual peak position, which moves towards $z=0$ as the coupling grows.

We note that the function in Eq.~\eqref{eq:si_pearcey_def} is familiar in paraxial optics, where a beam whose transverse profile is the Pearcey function auto-focuses and self-heals \cite{ring_optexpress_2012}.

\section{Energy bandwidth and duration of the compressed pulse}
\label{si:duration}

In the previous Sec.~\ref{si:pearcey}, we found the position at which the attosecond peak forms.
Here we obtain its width from the same diffraction integral, evaluated away from the
in-phase time at half of its maximum.

\textbf{Population spread.} The temporal width of the pulse is connected to the corresponding generated spectral width. In this case, the bandwidth is related to the standard deviation of the population number, which is defined as $\sigma_N^2=\langle N^2\rangle -\langle N \rangle^2$. Following from the sideband distribution of Eq.~\eqref{eq:si_cN} and using that $\langle N \rangle = 0$ due to symmetry (i.e., $P_{-N}=P_N$), and 
the Bessel identity $\sum_N N^2J_N^2(x)=x^2/2$ [Eq.~(8.536.2) in
Ref.~\citenum{gradshteyn2014table}],
\begin{equation}
\sigma_N^2= \sum_N N^2 J_N^2(2|g|)=2|g|^2,
\label{eq:si_second_moment}
\end{equation}
which, in this case, $\sigma_N$ equals the root-mean-square (RMS). Therefore, after a PINEM interaction, the energy spread
$\sigma_E=\hbar \omega \sigma_N = \sqrt2|g|\hbar\omega$, which is $9.9$~eV at $|g|=6$ and 1064~nm excitation wavelength (1.17~eV).

\begin{figure}[t]
\centering
\includegraphics[width=\linewidth]{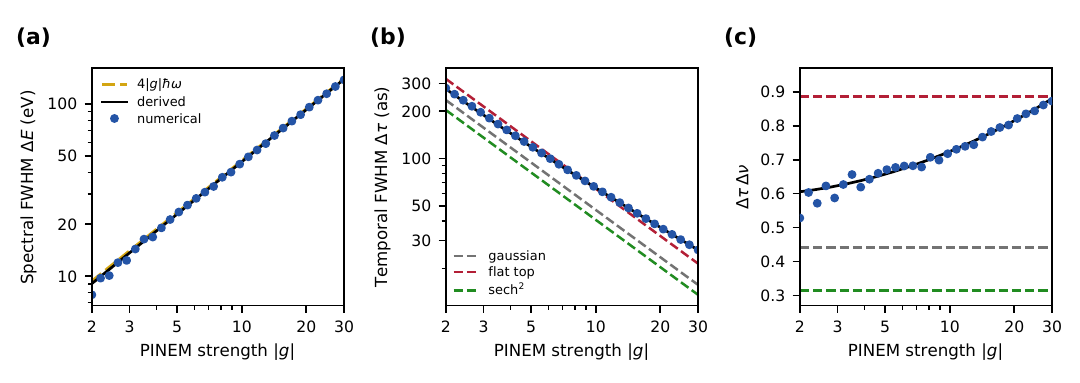}
\caption{\textbf{Widths of the compressed pulse against coupling
strength.} Circles are numerical in every panel, and each black solid line is
the closed form derived in this section. \textbf{(a)} Spectral full width at
half maximum of the energy comb, read from the outermost half-maximum crossings of the
spectrum broadened by the $0.7$~eV instrument function of Fig.~1(b) of the main
text. Black solid, Eq.~\eqref{eq:si_spec_fwhm}; and yellow dashed, is the classical estimate $4|g|\hbar\omega$ of Eq.~\eqref{eq:si_spec_classical}.
\textbf{(b)} Temporal full width at half maximum of the charge density at the
focus, from direct numerical propagation. Black solid, Eq.~\eqref{eq:si_fwhm};
broken coloured lines, the transform limits of Eq.~\eqref{eq:si_tl_general} for
a gaussian, a flat-top and a $\mathrm{sech}^2$ spectrum carrying the same RMS
bandwidth $\sigma_N=\sqrt2|g|$, distinguished by colour. \textbf{(c)}
Time--bandwidth product $\Delta\tau\,\Delta\nu$, the circles multiplying the two
numerical widths of panels (a) and (b).
Black solid, Eq.~\eqref{eq:si_tbp}; broken horizontal lines,
the transform-limited values $K$ of the three lineshapes of
Eq.~\eqref{eq:si_tl_general}, in the colours of the legend of panel (b). All
panels are for $10$~keV electrons and $1064$~nm driving wavelength.}
\label{figS_fwhm}
\end{figure}

In first approximation, the spectral width can then be obtained by computing the width of the spectral range created by the $P_N$ distribution. From the properties of the Bessel functions, the comb occupies $|N|\lesssim2|g|$, so the immediate estimate is the separation of the two classical extremes of the energy exchange,
\begin{equation}
\Delta E\simeq4|g|\hbar\omega ,
\label{eq:si_spec_classical}
\end{equation}
which is $28.0$~eV at the working point ($|g|=6$). Figure~\ref{figS_fwhm}(a) compares it
with the full width at half maximum of the numerically broadened comb (blue dots).
Equation~\eqref{eq:si_spec_classical} is exact only in the large values of $|g|$. As we shall see below, further corrections can be made of the order of $\mathcal O(|g|^{1/3})$.

\textbf{The duration of the pulse} from such a spectral width depends on the assumed spectral lineshape. For a spectral profile with a finite $\sigma_N$, it is convenient to characterize the lineshape by two
dimensionless numbers: (i) the ratio $f={\Delta\nu}/{\sigma_\nu}$
of the spectral full width at half maximum (FWHM), $\Delta\nu$, to its RMS width,
$\sigma_\nu$, and (ii) the transform-limited time--bandwidth product $K=\Delta\tau_{\rm TL}\Delta\nu$.
Using $\sigma_\nu=\sigma_N\omega/(2\pi)$ together with
Eq.~\eqref{eq:si_second_moment}, $\sigma_N=\sqrt{2}|g|$, gives
\begin{equation}
\Delta\tau_{\rm TL}
=
\frac{K}{\Delta\nu}
=
\frac{K}{f\sigma_\nu}
=
\frac{2\pi K}{f\,\sigma_N\omega}
=
\frac{1}{|g|\omega}
\frac{\sqrt{2}\pi K}{f}.
\label{eq:si_tl_general}
\end{equation}
The dependence of the minimum time--bandwidth product on pulse shape, including
the standard Gaussian and $\mathrm{sech}^2$ values, is discussed in
Sec.~1.1.3 of Ref.~\cite{diels2006ultrashort}. The values used here are
summarized in Table~\ref{tab:si_tl_lineshapes}. Note that in our first approach in Eq.~\eqref{eq:si_spec_classical}, we have $f = 2\sqrt{2}\approx 2.83$.

\begin{table}[h]
\caption{
Lineshape-dependent conversion between the spectral RMS width and the
transform-limited temporal FWHM. Here $f=\Delta\nu/\sigma_\nu$ and
$K=\Delta\tau_{\rm TL}\Delta\nu$. The Gaussian and $\mathrm{sech}^2$
time--bandwidth products are the standard transform-limited values discussed in
Ref.~\cite{diels2006ultrashort}; the flat-top value follows from the Fourier
transform of a rectangular spectral profile. The durations in the last column
are evaluated at $|g|=6$ using Eq.~\eqref{eq:si_tl_general}.
}
\label{tab:si_tl_lineshapes}
\begin{ruledtabular}
\begin{tabular}{lcccc}
Spectral profile
& $f$
& $K$
& $\sqrt{2}\pi K/f$
& $\Delta\tau_{\rm TL}$ at $|g|=6$ \\
\hline
Gaussian
& $2\sqrt{2\ln 2}=2.355$
& ${2\pi^{-1}\ln 2}=0.441$
& $\sqrt{\ln 2}=0.833$
& $78.4~\mathrm{as}$ \\
Flat top
& $2\sqrt{3}=3.464$
& $0.886$
& $1.136$
& $107.0~\mathrm{as}$ \\
$\mathrm{sech}^2$
& $1.944$
& $0.315$
& $0.720$
& $67.7~\mathrm{as}$ \\
\end{tabular}
\end{ruledtabular}
\end{table}

In particular, at $|g|=6$, these correspond to $78.4$~as, $107.0$~as, and $67.7$~as,
respectively, while numerically we obtained $101.6$~as. All three estimates therefore scale as $|g|^{-1}$ and are
independent of the electron energy. 
In Fig.~\ref{figS_fwhm}(b), we plot the FWHM that we obtain numerically and compare it with the lineshapes described above. Although the numbers for $|g|=6$ are very close to the ones obtained from propagating the electron charge density, they have an intrinsic $|g|^{-1}$ scaling that does not reproduce the simulated ones. 

Instead, we approach the problem from the diffraction integral Eq.~\eqref{eq:si_kirchhoff} evaluated at its maximum $\tau=\tau_f$.
The shape of the peak follows from the same integral in a neighborhood of that
time. Writing $\tau=\tau_f+\delta$ and $\tau'=\tau_f+\delta'$, so that the FWHM reads $\Delta\tau=2\delta$. Together with Eq.~\eqref{eq:si_inphase}, we have that $\sin(\omega\tau'-\phi)=-\cos(\omega\delta')$, and therefore, the exponent of the diffraction integral in Eq.~\eqref{eq:si_kirchhoff} separates into three pieces,
\begin{equation}
\frac{\omega^{2}(\tau-\tau')^{2}}{2u}-\sin(\omega\tau'-\phi)
=\frac{\omega^{2}\delta^{2}}{2u}
-\frac{\omega^{2}\delta\delta'}{u}
+\left[\frac{\omega^{2}\delta'^{2}}{2u}+\cos(\omega\delta')\right].
\label{eq:si_split_delta}
\end{equation}
The first term is free of the integration variable $\delta'$ and contributes a factor of
unit modulus, which vanishes when computing the density. Recalling the factor
$2\ii|g|$ that multiplies the whole exponent of
Eq.~\eqref{eq:si_kirchhoff}, the cosine is expanded about $\delta'=0$ (i.e., $\cos(x) \approx 1-x^2/2+x^4/24+\cdots$) and
truncated after its fourth-order term,
\begin{equation}
2|g|\left[\frac{\omega^{2}\delta'^{2}}{2u}+\cos(\omega\delta')\right]
\simeq
2|g|+|g|\,(\omega\delta')^{2}\,\frac{1-u}{u}
+\frac{|g|}{12}\,(\omega\delta')^{4}.
\label{eq:si_cos_expansion}
\end{equation}
Equation~\eqref{eq:si_svar} gives
$\omega\delta'=(12/|g|)^{1/4}s$, hence
$(\omega\delta')^{2}=2\sqrt3\,|g|^{-1/2}s^{2}$ and
$(\omega\delta')^{4}=(12/|g|)\,s^{4}$, so the three terms of
Eq.~\eqref{eq:si_cos_expansion} become a constant phase, $Bs^{2}$ with the $B$
of Eq.~\eqref{eq:si_pearcey_def}, and $s^{4}$ with unit coefficient. Note that the middle term of Eq.~\eqref{eq:si_split_delta} is linear in $s$
and carries the whole dependence on $\delta$,
\begin{equation}
2|g|\,\frac{\omega^{2}\delta\delta'}{u}=X s,
\qquad \text{with} \qquad
X=2\cdot12^{1/4}\,|g|^{3/4}\,\frac{\omega\delta}{u},
\label{eq:si_Xdef}
\end{equation}
so the exponent of Eq.~\eqref{eq:si_kirchhoff} is
$\ii(s^{4}+Bs^{2}-Xs)$ up to the constant phase. The integral is invariant under
$s\to-s$ together with $X\to-X$, which makes the peak symmetric in $\delta$, and lets us write
\begin{equation}
\psi(z,\tau_f+\delta)\propto
\mathcal P(X,B)=\int_{-\infty}^{\infty}ds\;
\ee^{\ii\left(s^{4}+Bs^{2}+Xs\right)} .
\label{eq:si_pearcey_2d}
\end{equation}
This is the canonical two-variable Pearcey function, of which
Eq.~\eqref{eq:si_pearcey_def} is the section $X=0$. Locating the focus and
resolving the pulse are therefore the same calculation along the two axes of the
maxima. At the focus $B=B^*$, and the density
$n\propto|\mathcal P(X,B^*)|^{2}$ is a function of $X$. Numerically,
$|\mathcal P(X,B^*)|^{2}$ decreases from its maximum at $X=0$, where
$|\mathcal P(0,B^*)|=2.6351$, and reaches half of that maximum at
$X_{1/2}=0.9832$, so that $\Delta X=2X_{1/2}=1.9663$.
Therefore, using Eq.~\eqref{eq:si_Xdef}, the FWHM yields
\begin{equation}
\Delta\tau=\frac{u}{2\cdot12^{1/4}\,|g|^{3/4}\,\omega}\;\Delta X .
\label{eq:si_fwhm_raw}
\end{equation}
Simplifying the prefactors, and using the definition of $u$ in Eq.~\eqref{eq:si_klystron} and evaluated at the focus $z_f$, we obtain
\begin{equation}
\Delta\tau=\frac{\Delta X}{24^{1/4}}\;\frac{1}{\omega\,(2|g|)^{3/4}}\;
\frac{z_f}{z_{f,\rm CL}},
\label{eq:si_fwhm_general}
\end{equation}
that expands as
\begin{equation}
\Delta\tau=\frac{0.888}{\omega\,(2|g|)^{3/4}}
\left[1+\frac{0.898}{\sqrt{2|g|}}+\frac{0.44}{2|g|}\right],
\qquad
\frac{\Delta X}{24^{1/4}}\approx0.888 .
\label{eq:si_fwhm}
\end{equation}
where we have used the result of Eq.~\eqref{eq:si_zf_practical} to simplify the result. The Eq.~\eqref{eq:si_fwhm} above reproduces the numerically propagated width
to better than $1.4\%$ over $2\le|g|\le30$, and at the working point of $|g|=6$ it gives
$100.9$~as against $101.6$~as [Fig.~\ref{figS_fwhm}(b)].

\textbf{Spectral width revisited.}
The spectral side admits a closed form of the same kind. The FWHM of the energy comb measures the edge of the comb at
$N=2|g|$ provided that the Lorentzian broadening $\ll 2|g|\hbar \omega$, the turning point of $J_N(2|g|)$, described by Eq.~(9.3.4) of
Ref.~\citenum{AS1972},
\begin{equation}
J_\nu\!\left(\nu+z\,\nu^{1/3}\right)
=2^{1/3}\nu^{-1/3}\,{\rm Ai}\!\left(-2^{1/3}z\right)+\mathcal O(\nu^{-1}),
\label{eq:si_airy_transition}
\end{equation}
which holds for $\nu\to\infty$ at fixed $z$. Applying it at $\nu=N$ and $z=(2|g|-N)/N^{1/3}$, and replacing $N^{1/3}$ by $(2|g|)^{1/3}$, the sideband populations become
\begin{equation}
P_N=J_N^2\!\left(2|g|\right)\simeq|g|^{-2/3}\,{\rm Ai}^2(t),
\qquad \text{with} \qquad
t=\frac{N-2|g|}{|g|^{1/3}} .
\label{eq:si_airy_comb}
\end{equation}
Note that the shape of $P_N$ near the
turning point is ${\rm Ai}^2(t)$, and therefore, the outermost maximum of $P_N$ sits at the first zero of ${\rm Ai}'$,
$t_0=-1.0188$ (see table 10.3 in Ref.~\cite{AS1972}), which is at $N=2|g|-1.0188\,|g|^{1/3}$, displaced inside the
classical turning point $N=2|g|$ by $\mathcal O(|g|^{1/3})$. 
For $t>t_0$ the function ${\rm Ai}^2$
decreases monotonically and reaches half of that maximum at $t_{1/2}=-0.0919$,
that is at $N=2|g|+t_{1/2}|g|^{1/3}$. Since $P_{-N}=P_N$ the distribution
is even in $N$, so the two outermost half-maximum crossings are separated by
$\Delta N=4|g|+2t_{1/2}|g|^{1/3}$ and
\begin{equation}
\Delta E=\hbar\omega\left[4|g|-0.184\,|g|^{1/3}\right],
\qquad \text{with} \qquad
0.184=2\,|t_{1/2}| ,
\label{eq:si_spec_fwhm}
\end{equation}
which at the working point ($|g|=6$) gives $27.6$~eV against a
numerical $26.3$~eV, with a $4.9\%$ error.

\textbf{Time-bandwidth product.}
From Eqs.~\eqref{eq:si_fwhm} and \eqref{eq:si_spec_fwhm}, and considering $\Delta\nu=\Delta E/2\pi\hbar$ we obtain the Time-bandwidth product to be
\begin{equation}
\Delta\tau\,\Delta\nu
=\frac{0.888}{2\pi}\;
\frac{4|g|-0.184\,|g|^{1/3}}{(2|g|)^{3/4}}
\left[1+\frac{0.898}{\sqrt{2|g|}}+\frac{0.44}{2|g|}\right],
\label{eq:si_tbp}
\end{equation}
in which $\omega$ has cancelled, as a dimensionless product requires, so the
result is a function of the coupling parameter. Keeping only the leading term of
each factor,
\begin{equation}
\Delta\tau\,\Delta\nu\simeq\frac{0.888}{\pi}\,(2|g|)^{1/4}
=0.283\,(2|g|)^{1/4},
\label{eq:si_tbp_lead}
\end{equation}
which shows the dependence on the coupling factor $g$. $\Delta\tau$ falls as $|g|^{-3/4}$ while the
bandwidth grows as $|g|$, so the product rises as $|g|^{1/4}$. Figure~\ref{figS_fwhm}(c) compares
Eq.~\eqref{eq:si_tbp} with the product formed from the simulation (blue dots), which
gives $\Delta\tau\,\Delta\nu=0.65$ at the working point ($|g|=6$), while 
Equation~\eqref{eq:si_tbp} gives $0.61$, $0.67$ and $0.88$ at $|g|=2$, $6$ and
$30$.

\section{Parallel paths: coherent and incoherent recombination}
\label{si:paths}

We derive here Eqs.~(10)--(13) of the main text. In the parallel geometry the
wave function is split among the arms, component $i$ carrying amplitude $p_i$
and acquiring its own coupling $g_i$, and the two limits differ only in what
happens once the components are brought back together. They add coherently
unless some other degree of freedom has become correlated with the arm index and
therefore holds a record of which near field a given component went through. 
The path density matrix $\Gamma_{ij}$ then quantifies how much of that record
survives: a diagonal $\Gamma$ means that the arm is fully registered and the
components cannot interfere, whereas a $\Gamma$ with non-vanishing out-of-diagonal elements means that no record
exists and they interfere completely. The formalism below
is the standard one for an interferometer of this kind
\cite{henke_pra_2025}. What is specific to PINEM is that the operation performed
in each arm is the displacement operator of Eq.~\eqref{eq:si_factorize}, and
that the coherences that survive are those of the energy ladder
\cite{kfir_sciadv_2021}. Writing the path-controlled operation as
$\hat U=\sum_i|i\rangle\langle i|\otimes\hat S_i$ and letting it act on the
uncorrelated initial state
$\Gamma\otimes\hat\rho^{(0)}=\sum_{i,j}\Gamma_{ij}|i\rangle\langle j|\otimes\hat\rho^{(0)}$,
the state after the interaction reads
\begin{equation}
\hat\rho
=
\sum_{i,j}\Gamma_{ij}\,|i\rangle\langle j|\otimes\hat S_i\hat\rho^{(0)}\hat S_j^\dagger ,
\label{eq:si_rho_after}
\end{equation}
which is Eq.~(9) of the main text.

For a given $\Gamma$, Eq.~\eqref{eq:si_rho_after} already fixes
the sideband populations,
\begin{equation}
P_N
=
\mathcal N_\coh^2\sum_{i,j}\Gamma_{ij}\,c_N(g_i)c_N^*(g_j),
\qquad
\mathcal N_\coh^{-2}=\sum_{i,j}\Gamma_{ij}\,J_0\!\left(2|g_i-g_j|\right),
\label{eq:si_PN_general}
\end{equation}
the normalization following from
$\sum_N c_N(g_i)c_N^*(g_j)=\langle\psi_j|\psi_i\rangle=J_0(2|g_i-g_j|)$ with
$|\psi_i\rangle=\hat S_i|0\rangle$. The two cases treated below are its limits,
a diagonal $\Gamma$, for which $\mathcal N_\coh=1$ because $J_0(0)=1$ and
${\rm Tr}\,\{\Gamma\}=1$, and $\Gamma_{ij}=p_ip_j^*$ that keeps every
cross term.

\textbf{Distinguishable paths.} If the record survives, the electron alone is
described by tracing Eq.~\eqref{eq:si_rho_after} over the path label. The path
states are orthogonal, ${\rm Tr}_{\rm path}\{|i\rangle\langle j|\}=\delta_{ij}$,
so the off-diagonal terms drop out and
\begin{equation}
\hat\rho^{(\rm i)}=\sum_i|p_i|^2\,\hat S_i\hat\rho^{(0)}\hat S_i^\dagger .
\label{eq:si_mixture}
\end{equation}
This map preserves the trace because each $\hat S_i$ is unitary and
$\sum_i|p_i|^2=1$. For an initially monochromatic electron,
$\rho^{(0)}_{nn'}=\delta_{n0}\delta_{n'0}$, it gives
$\rho^{(\rm i)}_{NN'}=\sum_i|p_i|^2c_N(g_i)c^*_{N'}(g_i)$ and therefore
\begin{equation}
P_N^{(\rm i)}=\sum_i|p_i|^2J_N^2\!\left(2|g_i|\right).
\label{eq:si_PN_mixture}
\end{equation}
which together with Eq.~\eqref{eq:si_mixture} gives Eqs.~(10) and (11) of the main text. We
stress that each term still carries the sideband coherences generated
within its own arm, since the incoherence refers only to the absence of
interference between arms.

\textbf{Erased which-path information.} If no record is left, the $i\neq j$
terms survive, the electron is described by
$\hat\rho^{(\rm c)}=\mathcal N_\coh^2\sum_{i,j}\Gamma_{ij}\,
\hat S_i\hat\rho^{(0)}\hat S_j^\dagger$, which is Eq.~(12) of the main text, and
its matrix elements read
\begin{equation}
\rho_{NN'}^{(\rm c)}
=
\mathcal N_\coh^2\sum_{i,j}\Gamma_{ij}\,c_N(g_i)c_{N'}^*(g_j).
\label{eq:si_coh}
\end{equation}
Its diagonal is the sideband distribution measured in a spectrum,
\begin{equation}
P_N^{(\rm c)}
=
\mathcal N_\coh^2\sum_{i,j}\Gamma_{ij}\,c_N(g_i)c_{N}^*(g_j)
=
\mathcal N_\coh^2\left|\sum_i p_i\,\ee^{\ii N \arg\{g_i\}}J_N\!\left(2|g_i|\right)\right|^2 ,
\label{eq:si_PN_coh}
\end{equation}
which is Eq.~(13) of the main text, where we have used $\Gamma_{ij}=p_ip_j^*$ to simplify the final expression. Comparing it with
Eq.~\eqref{eq:si_PN_mixture} isolates what the erasure adds, namely the cross
terms $i\neq j$: the arms are combined at the level of the amplitudes
$c_N(g_i)$ rather than of the populations $J_N^2(2|g_i|)$, so that the relative
optical phases $\arg\{g_i\}$ enter the spectrum.

The normalization $\mathcal N_\coh$ requires particular attention, and it is the one point at which the
parallel-path geometry differs from the sequential one. Tracing
Eq.~\eqref{eq:si_coh} over the sideband index gives
\begin{equation}
{\rm Tr}_N\{\hat\rho^{(\rm c)}\}
=
\mathcal N_\coh^2
\sum_{i,j}\Gamma_{ij}\,\langle0|\hat S_j^\dagger\hat S_i|0\rangle,
\qquad
\langle0|\hat S_j^\dagger\hat S_i|0\rangle=\sum_N c_N(g_i)c_N^*(g_j),
\label{eq:si_overlap}
\end{equation}
and the arm overlaps are in general not $\delta_{ij}$, because the states
$\hat S_i|0\rangle$ are not mutually orthogonal. Consequently
${\rm Tr}\,\{\Gamma\}=1$ does not by itself normalize the post-interaction
state, and we fix $\mathcal N_\coh$ by imposing
${\rm Tr}_N\{\hat\rho^{(\rm c)}\}=1$, so that
\begin{equation}
    \mathcal N_\coh^{-2}
=\sum_{i,j}\Gamma_{ij}\,\langle0|\hat S_j^\dagger\hat S_i|0\rangle
=\sum_{i,j}\Gamma_{ij}\,J_0\!\left(2|g_i-g_j|\right).
\end{equation}
For the two-arm example of Fig.~4 of the main text ($g_1=5$, $g_2=6\ee^{\ii\pi/4}$, and $p_1=p_2$) the unnormalized trace is $1.0093$, a $0.9\%$ correction, however now we ensure that $\sum_N P_N^{(\rm c)} = 1$.

\subsection{Propagation of the charge density}
\label{si:paths_propagation}

Free propagation multiplies each sideband by the quadratic phase of
Eq.~\eqref{eq:si_psi_prop}, so the density matrix of either case evolves as
\begin{equation}
\rho_{NN'}^{(\rm x)}(z)=\rho_{NN'}^{(\rm x)}(0)\,
\ee^{-2\pi\ii(N^{2}-N'^{2})z/z_T} ,
\label{eq:si_rho_prop}
\end{equation}
with ${\rm x}={\rm i}$ for the mixture of Eq.~\eqref{eq:si_mixture} and
${\rm x}={\rm c}$ for the which-path-erased case of Eq.~\eqref{eq:si_coh}.
The charge density in the co-moving frame is the diagonal element of this
matrix in the retarded-time representation,
$n^{(\rm x)}(z,\tau)=\sum_{N,N'}\rho_{NN'}^{(\rm x)}(z)\,\ee^{-\ii(N-N')\omega\tau}$,
so that only the difference of the two ladder indices reaches the retarded time. We
therefore label the terms by $m=N-N'$, the number of photon energies separating the two sidebands of a given
coherence, and sum over the index that remains, which we rename $N$. Each $m$
collects the elements lying $m$ energy levels above the diagonal and oscillates at
$m\omega$, so the density is a Fourier series in the optical period,
\begin{align}
n^{(\rm x)}(z,\tau)
=\sum_m \mathcal I_m^{(\rm x)}(z)\,\ee^{-\ii m\omega\tau}, 
\qquad
\text{with}
\qquad
\mathcal I_m^{(\rm x)}(z)
=\sum_N\rho_{N+m,N}^{(\rm x)}(z).
\label{eq:si_Im_contraction}
\end{align}
The term $m=0$ is the diagonal itself and carries the mean density, which
is also the spectrum of Eqs.~\eqref{eq:si_PN_mixture} and \eqref{eq:si_PN_coh}. Each arm enters through its own amplitudes $c_N(g_i)$, so the contraction
splits into the pair terms
\begin{align}
\mathcal I_m^{(ij)}(z)
=\ee^{\ii m\phi_i}\,\ee^{-2\pi\ii m^{2}z/z_T}
\sum_N J_{N+m}\!\left(2|g_i|\right)J_N\!\left(2|g_j|\right)
\ee^{\ii N\theta_{ij}} ,
\label{eq:si_Im_pair}
\end{align}
with $\phi_i=\arg\{g_i\}$ and $\theta_{ij}=\phi_i-\phi_j-4\pi m z/z_T$.
The two phases of Eq.~\eqref{eq:si_Im_pair} come from the same
rearrangement. Writing the element $\rho_{N+m,N}$ with the amplitudes of
Eq.~\eqref{eq:si_cN} and the propagation phase of Eq.~\eqref{eq:si_rho_prop}, the
quadratic exponent splits as $(N+m)^2-N^2=m^2+2mN$. The part $m^2$ does not involve
the summation index and comes out of the sum as $\ee^{-2\pi\ii m^2z/z_T}$, whereas
the part $2mN$ leaves $-4\pi m N z/z_T$ inside it. The amplitudes contribute
$\ee^{\ii(N+m)\phi_i}\ee^{-\ii N\phi_j}$, whose $N$-dependent part combines with
that remainder into the single angle $\theta_{ij}$, while $\ee^{\ii m\phi_i}$ joins
the prefactor. This is the same step as in Eqs.~\eqref{eq:si_quadratic_split} and
\eqref{eq:si_Im_presum} for a single arm, with the two arms now carrying different
couplings. These terms gather as
\begin{align}
\mathcal I_m^{(\rm i)}&=\sum_i |p_i|^{2}\,
\mathcal I_m^{(ii)},
\nonumber\\
\mathcal I_m^{(\rm c)}&=\mathcal N_\coh^{2}\sum_{i,j}
\Gamma_{ij}\,\mathcal I_m^{(ij)} .
\label{eq:si_Im_limits}
\end{align}
The mixture keeps only the diagonal elements of $\Gamma$, whereas the erased case retains all of its elements; the intermediate case is obtained by scaling the off-diagonal elements by a parameter $\mu\in[0,1]$.

The sum over $N$ in Eq.~\eqref{eq:si_Im_pair} is Graf's addition theorem,
Eq.~\eqref{eq:si_graf_applied}, now with the unequal arguments $2|g_i|$ and
$2|g_j|$, so that
\begin{align}
\mathcal I_m^{(ij)}(z)
&=\ee^{\ii m(\phi_i+\chi_{ij})}\,\ee^{-2\pi\ii m^{2}z/z_T}\,
J_m\!\left(w_{ij}\right),
\nonumber\\
w_{ij}&=2\left(|g_i|^{2}+|g_j|^{2}
-2|g_i||g_j|\cos\theta_{ij}\right)^{1/2},
\nonumber\\
\chi_{ij}&=\operatorname{atan2}\!\left(|g_j|\sin\theta_{ij},\;
|g_i|-|g_j|\cos\theta_{ij}\right) .
\label{eq:si_Im_graf}
\end{align}
Note that the four-quadrant form of $\chi_{ij}$ matters: writing only
$\tan\chi_{ij}$ leaves a $\pi$ ambiguity, which would reverse every odd harmonic.
Writing $x_m=2\pi m z/z_T$, two arms of the same coupling give
$w_{ii}=4|g_i|\,|\sin x_m|$ and the branch-independent relation
$\ee^{\ii m\chi_{ii}}J_m(w_{ii})=\ee^{\ii m(x_m-\pi/2)}J_m(4|g_i|\sin x_m)$. Since
$\ee^{-2\pi\ii m^{2}z/z_T}=\ee^{-\ii m x_m}$, the propagation phase cancels and
Eq.~\eqref{eq:si_Im_graf} returns Eq.~\eqref{eq:si_Im} arm by arm, so that the
mixture is a weighted sum of single-arm carpets, each of them in $\tau$. For
generally unequal couplings, $g_i\neq g_j$, that cancellation is incomplete,
because $\chi_{ij}$ depends on the propagation distance, and the cross terms carry
a phase that moves with $z$. This is what makes the interference pattern of the
which-path-erased case evolve along the beam while the mixture does not.

We verified Eqs.~\eqref{eq:si_Im_contraction}--\eqref{eq:si_Im_graf}
against a discrete Fourier transform of the numerically propagated density for the
two-arm example of Fig.~4 of the main text: the two agree to $9\times10^{-16}$ over
$m\le5$, and the contraction $\sum_N\rho_{N+m,N}$ reproduces the same coefficients
to $2\times10^{-16}$ in both the mixture and the erased case.

\section{Numerical implementation and validation}
\label{si:numerics}

All results were computed from the sideband representation directly, with no
approximation beyond the truncation of the ladder. Wave functions are evaluated as
$\psi(z,\tau)=\sum_{|N|\le N_{\max}}c_N(g)\ee^{-2\pi\ii N^2z/z_T}\ee^{-\ii N\omega\tau}$
with $N_{\max}=110$, for which $\sum_NP_N=1$ to better than $10^{-12}$ at the
couplings used here. Densities are normalized to the mean of the incident plane,
which equals ${\rm Tr}_N\{\rho\}$ and is conserved along $z$; we verify this to
$10^{-15}$ in every calculation, which also guards against the aliasing that a
retarded-time window spanning a non-integer number of optical periods would
introduce.

The implementation was validated against a set of independent identities:
unitarity of the amplitudes; $|\psi(0,\tau)|=1$ uniformly, i.e. the interaction
imprints pure phase; the closed-form harmonic law of Eq.~(4) of the main text
against a discrete Fourier transform of the propagated density, agreeing to
$3\times10^{-14}$; the second-moment identity Eq.~\eqref{eq:si_second_moment}; and
the equality of the statistical-mixture density with the weighted sum of individual
arm densities to machine precision. Focus positions obtained by direct
maximization of the propagated density agree with Eq.~\eqref{eq:si_zf_practical}
to $0.01\%$ at $|g|=6$ on a converged longitudinal grid, which is the figure
quoted in Methods; the coarser agreement apparent in Fig.~1(f) of the main text
is set by the resolution of the grid plotted there, not by the formula. An independent implementation in a second
language reproduces the reference values to the same precision.

As an external check, Eq.~\eqref{eq:si_zf_practical} was compared with published
numerical compression distances obtained independently of this work: $1.83$~mm
against the $1.8$~mm simulated in Ref.~\citenum{feist_nature_2015} at
$|g|=5.7$, and $2.78$~mm against the $2.75$~mm predicted in
Ref.~\citenum{priebe_natphoton_2017} at $|g|=3.95$, both for $120$~keV
electrons at $800$~nm. The classical estimate of Eq.~\eqref{eq:si_zcl} gives $1.40$ and $2.02$~mm in
these two cases, respectively, so it is the diffractive correction that brings them into agreement.

\newpage
\section{Supplementary figures}
\label{si:figures}

\begin{figure}[h]
\centering
\includegraphics[width=\linewidth]{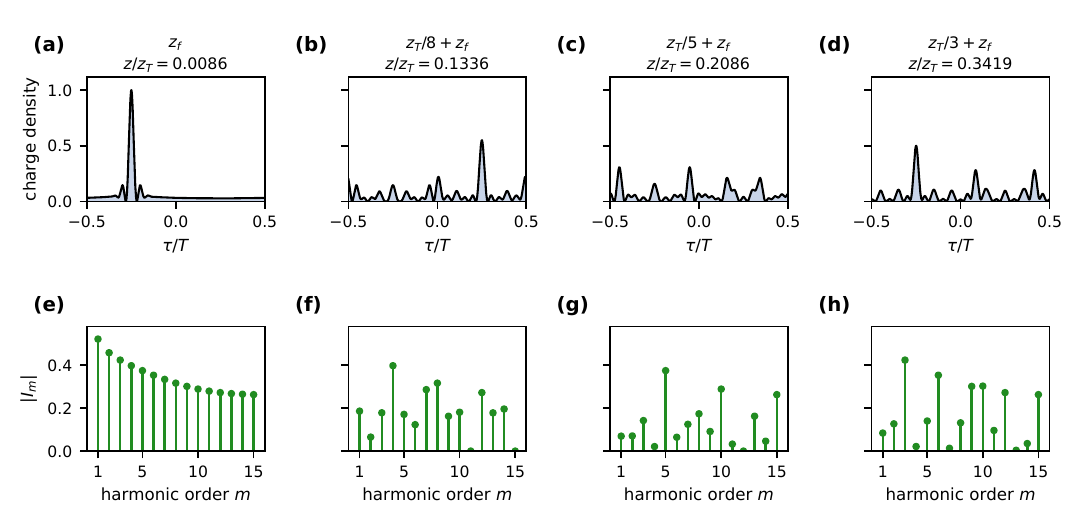}
\caption{\textbf{Slices of the Talbot maps of Fig.~2 of the
main text.} Columns run over the four planes marked by arrows in Fig.~2,
the focus $z_f$ and the fractional Talbot planes $z_T/8$, $z_T/5$ and $z_T/3$
measured from it, for $|g|=6$, $10$~keV electrons and $1064$~nm driving
wavelength. \textbf{(a--d)} Normalized charge density against retarded time $\tau$ for a given position $z/z_T$ given on top of each panel. \textbf{(e--h)} Harmonic
amplitudes $|\mathcal I_m|$ obtained from Eq.~\eqref{eq:si_Im}, which are a horizontal cut of Fig.~2(c), corresponding to the charge density of their top row.}
\label{figS_cuts}
\end{figure}

\begin{figure}[h]
\centering
\includegraphics[width=\linewidth]{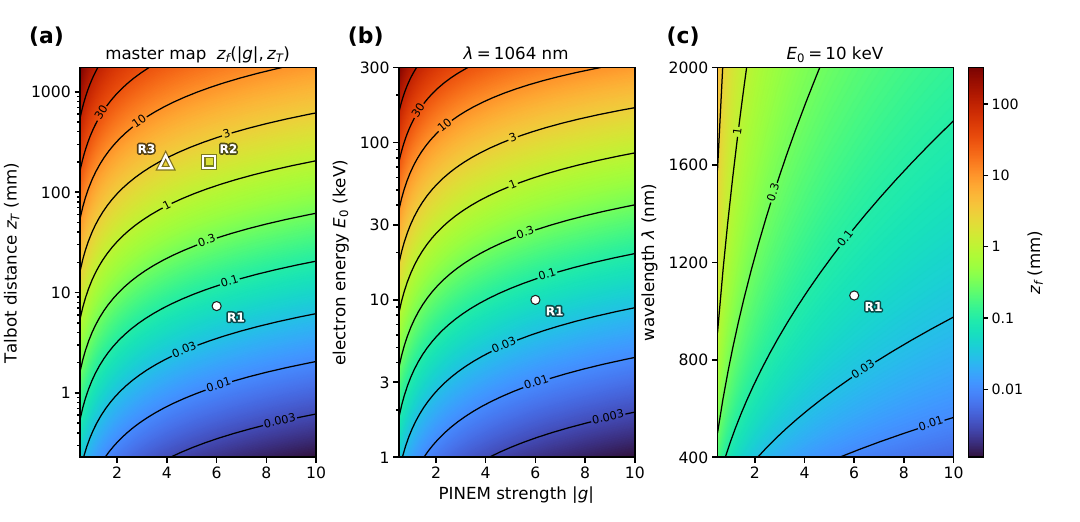}
\caption{\textbf{Design maps for the temporal focus.}
Focus distance $z_f$ from Eq.~\eqref{eq:si_zf_practical}, with labelled isocurves
of constant $z_f$ (in mm).
\textbf{(a)} Master map $z_f(|g|,z_T)$. Because the electron energy and the drive
wavelength enter only through the Talbot distance, this single map contains the
whole parameter dependence.
\textbf{(b)} $z_f(|g|,E_0)$ at fixed wavelength $\lambda=1064$~nm.
\textbf{(c)} $z_f(|g|,\lambda)$ at fixed electron energy $E_0=10$~keV.
Three reference points are marked in (a). The filled circle R1 is the working
point of this work, $10$~keV and $1064$~nm at $|g|=6$, where $z_T=7.33$~mm and
$z_f=63.0~\mu$m. The open square R2 and the open triangle R3 are the two
published cases against which Eq.~\eqref{eq:si_zf_practical} is checked in
Sec.~\ref{si:numerics}, $|g|=5.70$ in Ref.~\citenum{feist_nature_2015} and
$|g|=3.95$ in Ref.~\citenum{priebe_natphoton_2017}, both at $120$~keV and
$800$~nm, for which Eq.~\eqref{eq:si_zf_practical} gives $z_f=1.83$ and
$2.78$~mm. R2 and R3 share the same Talbot distance, $z_T=200.6$~mm, so they sit
at the same height and differ only in coupling.}
\label{figS1}
\end{figure}

\begin{figure}[H]
\centering
\includegraphics[width=\linewidth]{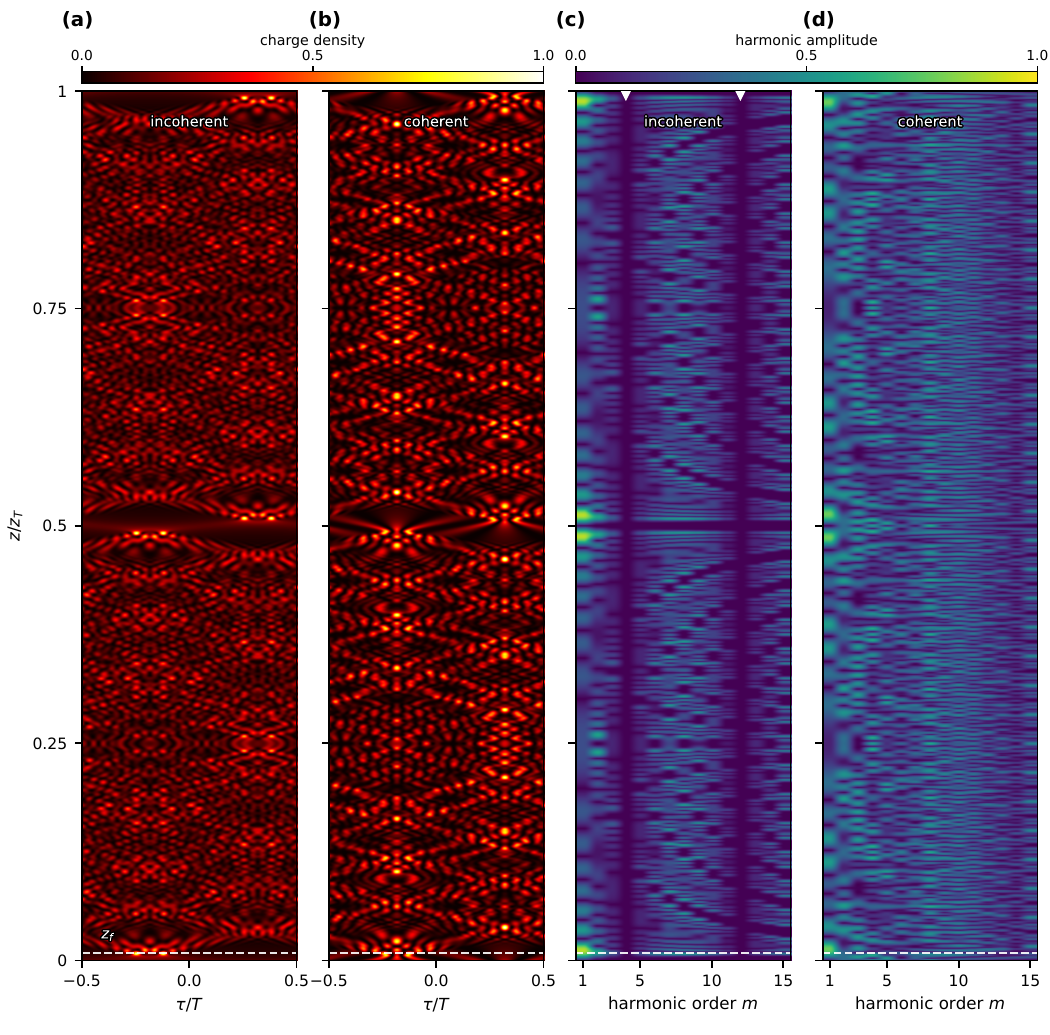}
\caption{\textbf{Talbot-scale propagation of the two-path
synthesis.} The two cases of Fig.~3(c,d) of the main text, $|g_1|=|g_2|=6$ with
$\arg\{g_2\}=\pi/4$, propagated over a full Talbot period instead of the
$0.05\,z_T$ window shown there, for $10$~keV electrons driven at
$1064$~nm ($\hbar\omega=1.17$~eV, $T=3.55$~fs). \textbf{(a,b)}
Charge density for the statistical mixture and for
the which-path-erased recombination, on a common
colour scale. The two arms carry equal weights,
$p_1=p_2=1/\sqrt2$, so that $\Gamma_{ij}=|p_i|^2\delta_{ij}=\delta_{ij}/2$ in the
mixture and $\Gamma_{ij}=p_ip_j^*=1/2$ for all $i,j$ in the which-path-erased
case. \textbf{(c,d)} Their harmonic
amplitudes $|\mathcal I_m|$ of Eq.~\eqref{eq:si_Im}, obtained by Fourier transforming each row of (a)
and (b) over one optical period and plotted against the same reference as
Fig.~2(c). The dashed line is the attosecond focus $z_f$ obtained from Eq.~\eqref{eq:si_zf_practical}.
The arm phase sets the time origin, so the mixture is the sum of two identical
carpets offset by $\arg\{g_2\}/\omega=T/8\approx0.444$~fs. Its harmonics therefore pick up the ordinary
two-copy factor and Eq.~\eqref{eq:si_Im} becomes $|\mathcal I_m|=|J_m(4|g|
\sin(2\pi m z/z_T))|\,|\cos(m\arg\{g_2\}/2)|$: the orders $m=4$ and $12$
(arrowheads), for which the offset is half a harmonic period, are empty at every
plane, whereas the coherent sum, whose cross term
is not a delayed copy of the single PINEM, carries them at $0.36$ and $0.24$. These
orders coincide with the sidebands that Eq.~\eqref{eq:si_PN_coh} removes from
the coherent spectrum of Fig.~3(b) because both cosines carry the half-argument
$\arg\{g_2\}/2$.}
\label{figS_talbot2arm}
\end{figure}

\begin{figure}[H]
\centering
\includegraphics[width=\linewidth]{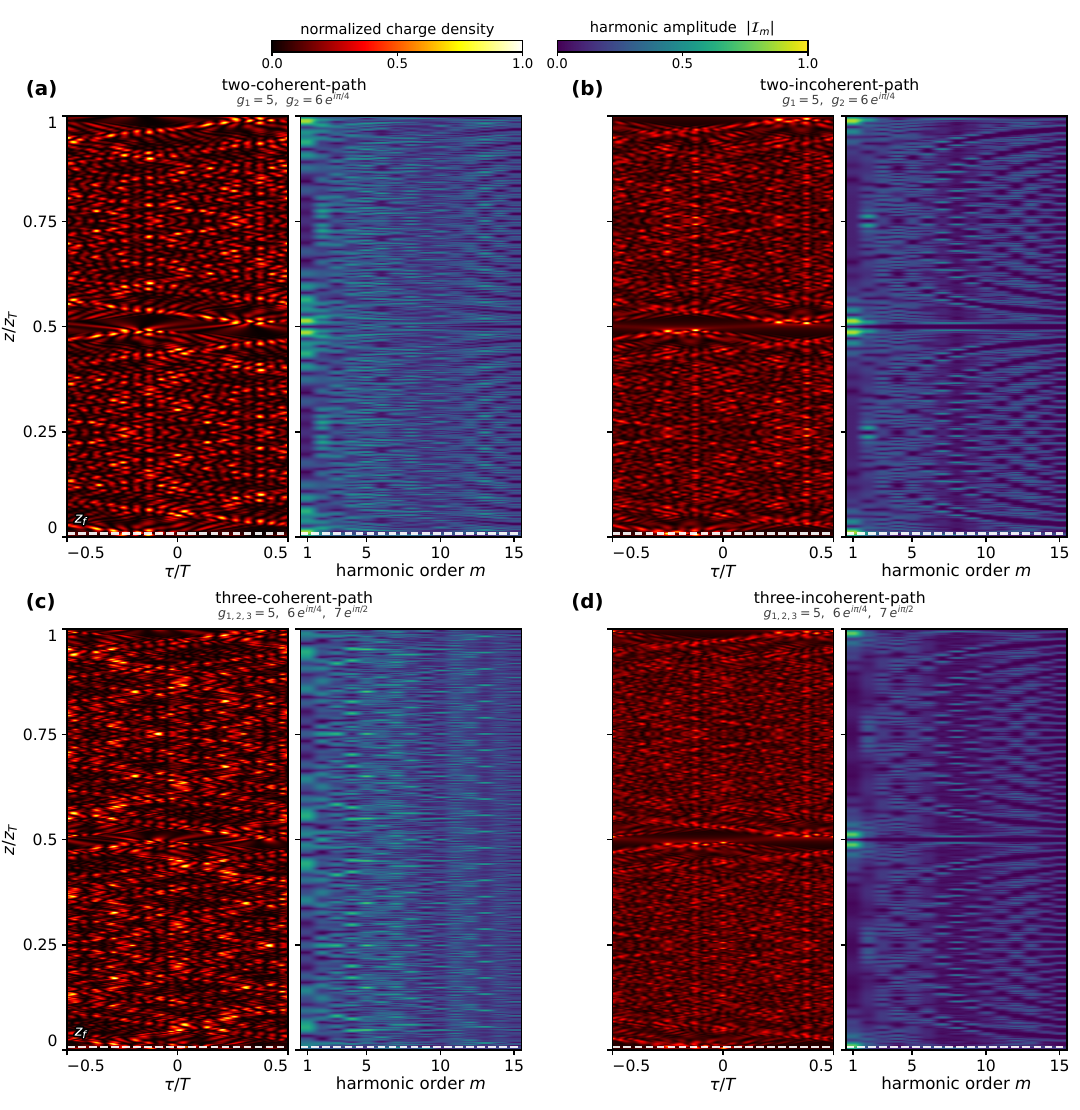}
\caption{\textbf{Talbot-scale propagation of the parallel-path channels.}
The four channels of Fig.~4 of the main text, propagated over one full Talbot
period. The arms are $g_1=5$,
$g_2=6\ee^{\ii\pi/4}$ and $g_3=7\ee^{\ii\pi/2}$ and all panels share the
propagation axis, for $10$~keV electrons driven at $1064$~nm
($\hbar\omega=1.17$~eV, $T=3.55$~fs). Each case shows the
charge density beside its harmonic amplitudes $|\mathcal I_m|$ of
Eq.~\eqref{eq:si_Im}, $m=1$ to $15$. The wave function is split evenly over the
$M$ arms, $p_i=1/\sqrt M$, so that $\Gamma_{ij}=|p_i|^2\delta_{ij}=\delta_{ij}/M$
in the statistical mixtures and $\Gamma_{ij}=p_ip_j^*=1/M$ in the
which-path-erased cases. The dashed line marks the attosecond focus of the
$|g|=6$ arm obtained from Eq.~\eqref{eq:si_zf_practical}. At the revival planes
$z/z_T=0.5$ and $1$ the statistical mixtures (b,d) return to a perfectly uniform
density, since each arm self-images to the incident beam and an incoherent sum of
uniform profiles is itself uniform, whereas the which-path-erased cases (a,c)
revive their inter-path interference pattern.}
\label{figS2}
\end{figure}

\begin{figure}[H]
\centering
\includegraphics[width=\linewidth]{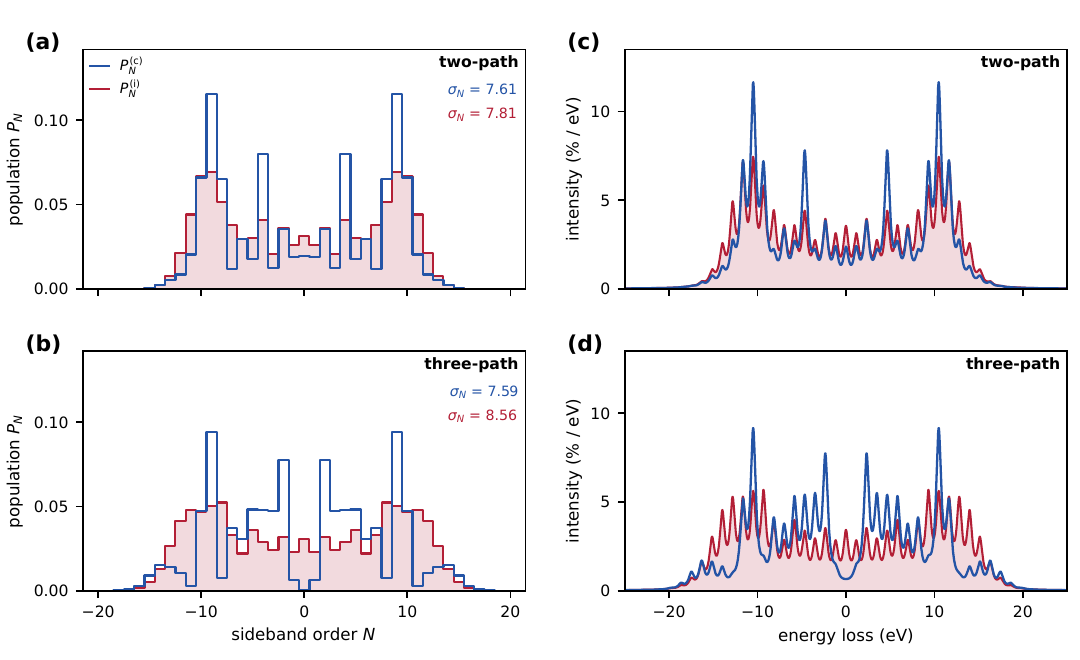}
\caption{\textbf{Energy-loss spectra of the parallel-path channels.}
Sideband populations (a,b) and the corresponding spectra after $0.7$~eV Lorentzian
instrumental broadening (c,d), for two- and three-path configurations, each in the
which-path-erased (coherent) and statistical-mixture cases (incoherent). The arms
carry different strengths and phases, $g_1=5$, $g_2=6\,\ee^{\ii\pi/4}$ and
$g_3=7\,\ee^{\ii\pi/2}$, the two-path case (a,c) using the first two and the
three-path case (b,d) all three. The wave function is split evenly over the $M$
arms, $p_i=1/\sqrt M$, so that the path density matrix is $\Gamma_{ij}=|p_i|^2\delta_{ij}=\delta_{ij}/M$ in the
incoherent case, Eq.~\eqref{eq:si_PN_mixture}, and has all $M^2$ elements equal,
$\Gamma_{ij}=p_ip_j^*=1/M$, in the coherent one, Eq.~\eqref{eq:si_PN_coh}. Those
off-diagonal elements let the arms add as amplitudes, so that every pair of arms
contributes an interference term $\cos(N\Delta\phi_{ij})$ to the sideband
envelope, with $\Delta\phi_{ij}=\arg\{g_j\}-\arg\{g_i\}$ and a period of
$2\pi/\Delta\phi_{ij}$ sidebands. Here $\arg\{g_1\}=0$ and $\arg\{g_2\}=\pi/4$, so
the first pair beats every eight sidebands, or $9.3$~eV, while $\arg\{g_3\}=\pi/2$
adds a four-sideband beat, $4.7$~eV, in the three-path case, using the 1064~nm excitation wavelength (1.17~eV). The root-mean-square widths are shown in each panel. Panels (a,b) show the populations themselves, $\sum_N P_N=1$,
and (c,d) the broadened intensity in per cent per eV.}
\label{figS3}
\end{figure}

\clearpage
\bibliographystyle{apsrev4-2}
\bibliography{refs_paper1}